\documentclass[preprint]{aastex}
\usepackage{amsmath}
\newcommand{\h}{{\rm H}}

\newcommand{\xh}{x({\rm H})}
\newcommand{\nh}{n_{\rm H}}
\newcommand{\Nh}{N_{\rm H}}
\newcommand{\hp}{{\rm H}^+}
\newcommand{\xhp}{x({\rm H}^+)}
\newcommand{\nhp}{n({\rm H}^+)}
\newcommand{\hm}{{\rm H}_2}

\newcommand{\hmp}{{\rm H}_{2}^{+}}

\newcommand{\hthreep}{{\rm H}_{3}^{+}}

\newcommand{\HHHOp}{{\rm H}_{3}{\rm O}^{+}}

\newcommand{\he}{{\rm He}}

\newcommand{\hep}{{\rm He}^+}
\newcommand{\xhep}{x({\rm He}^+)}

\newcommand{\xel}{x_{\rm e}}
\newcommand{\nel}{n_{\rm e}}
\newcommand{\ha}{{\rm A}}
\newcommand{\xha}{x({\rm A})}
\newcommand{\hap}{{\rm A}^+}
\newcommand{\xhap}{x({\rm A}^+)}
\newcommand{\happ}{{\rm A}^{2+}}
\newcommand{\hatwop}{{\rm A}^{2+}}

\newcommand{\nep}{{\rm Ne}^+}

\newcommand{\netwop}{{\rm Ne}^{2+}}

\newcommand{\htwoo}{{\rm H}_{2}{\rm O}}
\newcommand{\hcop}{{\rm HCO}^+}
\newcommand{\LX}{L_{\rm X}}

\newcommand{\pcc}{{\rm cm}^{-3}}

\newcommand{\psqcm}{{\rm cm}^{-2}}
\newcommand{\persqcm}{\rm \,cm^{-2}}
\newcommand{\gpersqcm}{\rm \,g\,cm^{-2}}

\newcommand{\ps}{{\rm \,s}^{-1}}
\newcommand{\ccps}{{\rm cm}^{3} {\rm s}^{-1}}

\newcommand{\ergps}{{\rm erg}\,{\rm s}^{-1}}
\newcommand{\bcen}{\begin{center}}
\newcommand{\ecen}{\end{center}}
\newcommand{\be}{\begin{equation}}
\newcommand{\ee}{\end{equation}}
\newcommand{\bdis}{\begin{displaymath}}
\newcommand{\edis}{\end{displaymath}}
\newcommand{\Td}{T_{\rm d}}
\newcommand{\Tg}{T_{\rm g}}
\newcommand{\Msun}{M_{\odot}}
\newcommand{\Rsun}{R_{\odot}}
\newcommand{\Mst}{M_{\ast}}

\newcommand{\Rst}{R_{\ast}}

\long\def\symbolfootnote[#1]#2{\begingroup%
\def\thefootnote{\fnsymbol{footnote}}\footnote[#1]{#2}\endgroup}

\def\ra{{\rightarrow}}

\begin{document}

\title{X-ray Ionization of Heavy Elements Applied to Protoplanetary Disks}

\author{M\'{a}t\'{e}  \'{A}d\'{a}mkovics\altaffilmark{1}, 
Alfred E. Glassgold\altaffilmark{1}, Rowin Meijerink\altaffilmark{2}}

\altaffiltext{1}{Astronomy Department, University of California,
Berkeley, CA 94720-3411, USA} 
\altaffiltext{2}{Leiden Observatory, Leiden University, P.O. Box 9500, 2300 RA Leiden,
The Netherlands}

\begin{abstract}

The consequences of the Auger effect on the population of heavy element ions are analyzed for the case of relatively cool gas irradiated by keV X-rays, with intended applications to the accretion disks of young stellar objects. Highly charged ions are rapidly reduced to the doubly-charged state in neutral gas, so the aim here is to derive the production rates for these singly- and doubly-charged ions and to specify their transformation by recombination, charge transfer, and molecular reactions. The theory is illustrated by calculations of the abundances of eleven of the most cosmically abundant heavy elements in a model of a typical T Tauri star disk. The physical properties of the gas are determined with an X-ray irradiated thermal-chemical model, which shows that the disk atmosphere consists of a hot atmosphere overlaying the mainly cool body of the disk. There is a warm transition layer where hydrogen, carbon and oxygen are transformed from atomic to molecular species and the ionization drops by several orders of magnitude. The doubly charged ions are then largely confined to the hot outer layers of the disk. The nature of the dominant ions below the transition depends sensitively on the poorly constrained abundances of the heavy elements. Observational consequences and connections with the active layers of the magneto-rotational instability are briefly discussed.
  
\end{abstract}

\section{Introduction} 

It is well established that young stellar objects (YSOs) are strong emitters of X-rays and that they have the potential to affect the physical and chemical properties of circumstellar material (e.g., Feigelson \& Montmerle 1999; Glassgold, Feigelson \& Montmerle 2000). Most attention has been given to their effect on the inner regions of the accretion disks of low-mass YSOs (e.g., Glassgold, Najita \& Igea 2004 [GNI04]; Walsh et al.~2010; Shang et al.~2002, 2009 for applications to jets). The YSO ionizing spectra range from relatively soft extreme ultraviolet (EUV) to hard keV photons. The latter are able to penetrate the surface layers of disks and play a  role in determining the physical, chemical and ionization properties of the observable regions of protoplanetary disks.

The hard keV X-rays are mainly absorbed by photoionization from the K and L shells of heavy atoms, resulting in the production of multiply-charged ions due to the Auger effect. In atomic regions, these highly-charged ions are likely to be destroyed by charge exchange with atomic H, although this process is slow for some singly- and doubly-charged ions. In molecular regions, the ions may be rapidly destroyed by reactions with molecular hydrogen and other species, but there are important differences between singly- and doubly-charged ions in this respect. The significance of the reactions depend on whether the heavy atom is chemically active or not. The role of X-ray production of multiply charged atomics ions was first discussed almost forty years ago for the interstellar medium (e.g., Weisheit and Dalgrno 1972; Weisheit 1973, 1974; Watson and Kunz 1975; Watson 1976).

We develop a systematic theory of the ion abundances of eleven of the most abundant heavy atoms. The ions are produced by X-rays and destroyed by charge exchange, electronic recombination, and a variety of molecular reactions. We illustrate the theory using the D'Alessio et al.~(1999) model of a typical T Tauri star disk. We show that the transition from an atomic to a molecular region that occurs in the upper atmosphere is accompanied by a significant drop in the ionization level (the electron fraction $\xel$) and by a large increase in the abundance of molecular ions. Although these ions play an important role in regulating the ionization level, singly-charged heavy atomic ions can dominate the ionization below the transition region. In this respect, our work supports the classic paper by Oppenheimer \& Dalgarno (1974) on the cosmic ray ionization of dense interstellar
clouds.

In the next section (Section 2) we derive formulae for the ionization rates of the ions of each element based on consideration of the three main ionizations processes: photoionization, the Auger Effect, and the production of secondary electrons in electronic collisions. In Section 3, we describe the reactions that determine the ionization state populations for each element. In Section 4 we present results using our earlier thermal-chemical model (GNI04); these include ionization rates and ion populations as well as the temperature and ionization structure of the model disk atmosphere extending down to surface densities of several $\gpersqcm $. The paper concludes with a general discussion of the results that includes some perspectives on the diagnostic potential of spectral lines of atomic ions. 

\section{Theory} 

\subsection{Physical Processes} 

Our goal is to calculate the rate  at which stellar X-rays ionize heavy atoms in protoplanetary disks, specifically in the upper layers of the disks of evolved low-mass YSOs  in the T Tauri phase. We assume that, at this stage, the gas in these regions has been depleted in heavy elements, at least to the extent  seen in diffuse interstellar clouds (e.g., Jenkins 2009). These regions are also expected to have experienced grain growth and settling (e.g.,  Furlan et al.~2006), with the grain surface area per unit hydrogen nucleus reduced by two or three orders of magnitude, e.g., to $\sim$10$^{-23}$\,cm$^2$ for a reduction factor of 100. Under these circumstances, the primary interaction of stellar keV X-rays with the disk atmosphere is photoionization of gas-phase heavy atoms  and singly-charged ions. The primary interaction produces a fast photo-electron plus several more fast electrons generated by the accompanying fluorescence  and the Auger Effect. These primary electrons then produce many more secondary electrons by collisionally ionizing H, $\hm$ and He. Thus X-ray ionization proceeds in two ways, {\it direct} ionization immediately following absorption (photoionization, fluorescence, and the Auger Effect) and {\it secondary} ionization due to the secondary electrons resulting from the direct ionization.

Fluorescence is usually a minority mode compared to the Auger Effect, except for the heaviest cosmically abundant elements. Its probability is $\sim 0.1$ for atomic numbers $Z \leq 20$ and reaches 0.35 for $Z = 26$ (Kaastra \& Mewe 1993).  Most of the fluorescent emission following K-shell absorption is at optical or UV wavelengths (Shapiro \& Bahcall 1981), so some of these photons will escape the disk and some will be absorbed locally and contribute to heating and ionization. For the case of iron about 30\% of the fluorescent photons are X-rays that can contribute to local ionization and heating. However, iron is likely to be heavily depleted, and fluorescence can be safely ignored in calculating the direct ionization rate per hydrogen nucleus, which is an abundance-weighted sum over all elements (Eq.~\ref{tot_dir} below). Our neglect of fluorescence, with a probability of $\sim$0.1, leads to an over-estimate of the direct ionization rate by $\sim$10\% (atomic data reviewed by Kallman \& Palmeri (2007), and references therein).

The mean number of Auger electrons is given by Kaastra \& Mewe (1993) for each heavy atomic ion on the basis of single-particle quantum calculations. It depends on whether photoionization occurs from the K, L, or M shell. For example, for the K shell, the mean number of Auger electrons is $\sim$1 for $Z \leq 10$, and it increases rapidly with $Z$ and reaches  $\sim$5.5 for iron. Not only do the Auger electrons carry away a significant amount of energy, typically 90\% of the threshold energy, but the Auger Effect leads to the production of multiply-charged ions. These considerations play an important role in the calculation of the ionization rates of heavy atomic ions.

The fast photo- and Auger electrons from direct ionization generate a cascade of secondary electrons in collisions with both the ambient thermal electrons and with H, $\hm$ and He; the latter are dominated by relatively small energy losses, tens of eV, as measured in the laboratory (Opal, Peterson \& Beaty 1971; Opal, Beaty, \& Peterson 1972). Electron energy-loss has a long history, and we refer to Dalgarno, Yan \& Liu (1999) for a  comprehensive study of H, $\hm$ and He mixtures and to recent calculations for H and He mixtures by Furlanetto \& Stoever (2010). 
The energy expended to make an ion pair is a constant $W\simeq37$\,eV for X-ray energies $> 100$\,eV and electron fractions $\xel < 0.1$ (Dalgarno et al. 1999). We give a simplified treatment of secondary electron ionization based on this quantity and on ideas of Maloney, Hollenbach \& Tielens (1996), as has also been done by Meijerink \& Spaans (2005). 
 
\subsection{Ionization Rates} 

On the basis of the above discussion, we adopt the terminology, direct and secondary ionization of heavy atoms or ions. By {\it direct} we mean ionization by the photoelectric effect and by the accompanying Auger electrons, typically at keV-energies. By {\it secondary} we mean electronic ionization by the secondary electrons produced by direct ionization. These three processes are described by the the cross section for the  absorption of a photon of energy $E$ by atom A, $\sigma_{\ha}(E)$, and the cross section for electronic ionization of atom A by an electron with energy $e$, $\sigma_{\ha, \,\rm ion} (e)$. We use Verner \& Yakolev (1995) for X-ray absorption cross sections,  Kaastra \& Mewe (1993) for Auger electron probabilities and standard sources like Tarawa \& Kato (1987), Mattioli et al.~(2007) and
Voronov (1997) for electronic ionization. 

If $F(E)$ is the local spectral distribution of the X-ray number flux at a distance $r$ from the stellar X-ray source, the primary photoionization rate of atom A is,
\be
\label{zetapriA}
\zeta_{\ha,\, \rm phel} = \int_{E_0}^{\infty}\, dE\,
F(E) \, \sigma_{\ha}(E).
\ee  
In practice, as discussed in the next section, $F(E)$ decreases with increasing $r$ because of inverse square dilution and attenuation along the line of sight. For large vertical column densities, this decrease is altered by Compton scattering (Igea \& Glassgold 1999). Before entering the disk, the X-rays may have experienced attenuation and scattering close to the source. As in the past, we use the simple expedient of removing all of the EUV and the soft X-ray photons up to some energy $E_0$, of order 200\,eV (the value we use in applications). When we sum this result over all elements and multiply by their abundances, we obtain the primary photoionization rate per unit volume per H nucleus, \be
\label{totzetapri}
\zeta_{\rm phel} = \int_{E_0}^{\infty}\, dE\,
F(E) \, \bar{\sigma}(E),
\ee
where
\be
\label{avxsec}
\bar{\sigma}(E) = \sum_{\ha}\, x_{\ha} \, \sigma_{\ha}(E)
\ee  
is the average photoionization cross section, and the abundances $x_{\ha}$ are normalized relative to hydrogen. The direct ionization rate, which combines electron production by the photoionization and the Auger Effect, is
\be
\label{zetadirA}
\zeta_{\ha,\, \rm dir} = \int_{E_0}^{\infty}\, dE\,
F(E) \, \sigma_{\ha}(E)\, .
\ee  
The direct ionization rate per H nucleus is then the sum of these integrals over all target atoms and ions,
\be
\label{tot_dir}
\zeta_{\rm dir} = \sum_{\ha}\, x_{\ha} \, \zeta_{\ha,\, \rm dir}\,. 
\ee

The rate at which the secondary electrons ionize a heavy atom is determined by their absolute number distribution distribution $n_{\rm sec} (e)$ as a function of the electron energy, denoted by $e$ in the following integral, 
\be
\label{zetaAsec}
\zeta_{\ha, \, \rm sec} = 
\int \, de \, n_{\rm sec} (e) \, v(e) \, \sigma_{\ha, {\rm ion}} (e), 
\ee 
where $v(e)$ is the speed of a secondary electron and $\sigma_{\ha, {\rm ion}}(e)$ is the cross section for the production of $\hap$ from $\ha$ in an electronic collision. In accord with experiments and theory on electronic ionization (e.g., Tarawa \& Kato 1987), we can ignore multiple ionization and express the production rate of secondary electrons per H nucleus as,
\be
\label{totzetasec}
\zeta_{\rm sec} = \sum_{\ha}\, x_{\ha} \, \zeta_{\ha, \, \rm sec}\,. 
\ee 

\subsection{Practical Implementation} 

Equations~\ref{tot_dir} and \ref{totzetasec} lead to the ionization rate per H nucleus, 
\be
\label{totrate} 
\zeta = \zeta_{\rm dir} + \zeta_{\rm sec}. 
\ee
The second term dominates the first by more than an order of magnitude and we use the approximation $\zeta \simeq \zeta_{\rm sec}$ in the rest of this paper. The calculation of $\zeta_{\rm sec}$ requires a knowledge of the secondary electron distribution according to 
Eq.~\ref{zetaAsec}. However, this can be a formidable task in a large thermal-chemical disk program, so rather 
than evaluating 
Eqs.~\ref{zetaAsec} and \ref{totzetasec}, we adopt the reliable semi-empirical formula,
\be
\label{approxtotzeta}
\zeta_{\rm sec} \simeq   
\int_{E_0}^{\infty}\, dE\, F(E) \, \bar{\sigma}(E)\, \frac{E}{W},
\ee  
based on the average energy to make an ion pair in cosmic gas $W$, introduced in Sec.~2.1. $W=37$\,eV is a good approximation in mainly neutral regions where $\xel< 0.1$ (Dalgarno, Yan \& Liu 1999). 

However, we still need the secondary ionization rates for the individual heavy elements, and here we follow Maloney, Hollenbach \& Tielens (1996) and develop an approximate treatment that makes use of the relative rather than the absolute secondary energy distribution,
\be
f_{\rm sec}(e) \equiv n_{\rm sec} (e)/n_{\rm sec}, 
\ee
where $n_{\rm sec}$ is the total density of secondary electrons,
\be
n_{\rm sec} = \int_0^{\infty}\, de \, n_{\rm sec} (e).
\ee
The absolute secondary electron distribution $n_{\rm sec} (e)$ appears in the integral in Eq.~\ref{zetaAsec} for the secondary ionization rate for atom A and thus also in Eq.~\ref{totzetasec}  for the secondary rate per H nucleus. Because electronic ionization cross sections are all of the same order of magnitude, only the most abundant terms in Eq.~\ref{totzetasec} are important,
\be
\zeta_{\rm sec} \simeq \zeta_{\h, \rm sec} + 
                        x_{\he} \, \zeta_{\he, \rm sec}.
\ee
In the first term we have combined the atomic and molecular hydrogen contributions since the cross section for $\hm$ is approximately twice that for atomic H. The second term is itself smaller than the first by an order of magnitude since it contains the abundance of He. We use this approximation to determine the secondary ionization rate of A, which is a well-defined fraction of the total secondary ionization rate according to Eq.~\ref{totzetasec};
\be
\frac{\zeta_{\ha, \, \rm sec}}{\zeta_{\rm sec}} =
\frac{\zeta_{\ha, \rm sec}}
{\sum_{\ha}\, x_{\ha} \, \zeta_{\ha, \, \rm sec}}	 
\simeq \frac{\zeta_{\ha, \rm sec}}
{\zeta_{\h, \rm sec} + x_{\he}\zeta_{\he, \rm sec}}.  
\ee
Using our basic approximation, $\zeta \simeq  \zeta_{\rm sec}$, this becomes
\be
\zeta_{\ha, \, \rm sec} \simeq 
\frac{\zeta_{\ha, \rm sec}}
{\zeta_{\h, \rm sec} + x_{\he}\zeta_{\he, \rm sec}} \, \zeta.
\ee
Finally, if He collisions are ignored, this last equation reduces to
\be
\label{secA}
\zeta_{\ha, \rm sec} \simeq  
\frac{\int \, de \, f_{\rm sec} (e)\, \bar{v}(e) \, \sigma_{\ha, {\rm ion}} (e)}{\int \, de \, f_
{\rm sec}(e)  \, \bar{v}(e) \, \sigma_{\h, {\rm ion}} (e) 
} \, \zeta
= \frac{<\sigma_{\ha, {\rm ion}}>}{<\sigma_{\h, {\rm ion}}>}\, \zeta,
\ee
where $ < \ldots > $ indicates an average over $f_{\rm sec} (e)$. Now following Maloney et al.~(1996) closely, we replace the ratio of the two cross section integrals by the ratio of the cross sections at some typical energy, again avoiding the calculation of the relative secondary electron distribution. We choose the characteristic energy to be that of the peak of the cross section. Table 1 gives these ratios for A and $\hap$ obtained from Tarawa \& Kato (1987).

\begin{center}
Table 1 \\
\begin{tabular}{ccccc}    
\multicolumn{5}{c}{Peak Electronic Ionization Cross Sections$^\mathrm{a}$} \\
\hline
\hline
Element & $\sigma(\ha)$	& $\sigma(\hap)$ & $\sigma(\ha)/\sigma_{\h}$ 	&
$\sigma(\hap)/\sigma_{\h}$   \\        
\hline
H	& 0.67	& \dots	& 1.0	& \dots	\\
C	& 2.3	& 0.55	& 3.4 	& 0.82	\\
N	& 1.5	& 0.50	& 2.2	& 0.75	\\
O	& 1.5	& 0.45	& 2.2	& 0.67	\\
Ne	& 0.75	& 0.32	& 1.1 	& 0.48	\\
Na	& 4.88	& 0.27	& 7.3	& 0.40	\\
Mg	& 5.3	& 0.43	& 7.91	& 0.64	\\
Si	& 6.69	& 1.56	& 10.0	& 2.3	\\
S	& 4.50  & 1.39	& 6.7	& 2.1	\\
Ar	& 2.5	& 1.2	& 3.7	& 1.8	\\
K	& 9.0	& 1.00	& 13	& 1.5	\\
Fe	& 5.34	& 1.08	& 8.0	& 1.6	\\
\hline

\multicolumn{5}{l}{$^\mathrm{a}$ Cross section units: $10^{-16}$\,cm$^2$.}\\
\end{tabular}
\end{center}

\noindent It is difficult to evaluate the accuracy of the approximate formula, Eq.~\ref{secA}, in the absence of actual calculations of the secondary electron distribution.

The discussion above refers to ``A'' as a general heavy atom where the notation suggests {\it neutral} atoms. However, the theory applies generally to any ionization stage. In practice,  ions beyond $\hatwop$ have negligible abundances because of rapid charge exchange with atomic and molecular hydrogen and other species, so the general formulae in Eqs.~\ref{zetadirA} and \ref{zetaAsec} refer to both A and $\hap$, and general expressions for rates like Eqs.~\ref{tot_dir} and \ref{totzetasec} involve sums over both A and $\hap$. Similarly, Eq.~\ref{secA} is supplemented by
\be
\label{secA+}
\zeta_{\hap, \, \rm sec} \simeq 
\frac{<\sigma_{\hap, \, {\rm ion}}>}{<\sigma_{\h, {\rm ion}}>}\, \zeta.
\ee
On this basis we can write the volumetric production rate of $\hap$ and $\hatwop$ by secondary electrons as,
\be
\label{prodAsec}
P_{\ha^+, \, \rm sec} =  \nh \, x_{\ha} \, \zeta_{\ha, \, \rm sec} 
\hspace{0.5in}
P_{\hatwop, \, \rm sec} =  \nh  \, x_{\ha^+} \, \zeta_{\hap, \, \rm sec} 
\ee 
where $x_{\ha}$ and $x_{\hap}$ are the abundances of A and $\hap$ and $\nh$ is the volume density of hydrogen nuclei.

\subsection{Direct Rates by Shell} 

When developing rate equations to describe the ionization of heavy atoms and ions, the ionization state of the product and the number of electrons released must be considered. Our reaction network treats only singly and doubly charged ions as chemically distinct species because atomic ions with higher charge are rapidly destroyed by charge exchange and other reactions. The results of secondary ionization are straightforward and are given by Eq.~\ref{prodAsec}. However, due to the Auger Effect, direct ionization can produce a variety of charge states depending on the X-ray energy, i.e., on whether the K, L, or M threshold edges are reached. The edge energies increase with atomic number, e.g., for iron they occur at $\sim$50-90\,eV, $\sim$700-850\,eV and $\sim$7100\,eV for the M, L and K shells, respectively. When an inner shell is vacated, multiple Auger electrons may be released. Kaastra \& Mewe (1993) calculate the probability distribution for the number of electrons released in this process for each element and vacated inner shell.  We approximate these probabilities, which are less than a few percent and greater than 90\%, by zero and 100\% likelihood for the production of either one electron or multiple electrons. Our estimates are essentially the same as the calculated values for the light elements through Ne, where the probability for one Auger electron is very close to 100\%. Starting with Na, there is a significant probability for two or more Auger electrons, but the sum of the probabilities for two or more Auger electrons still adds up to near 100\%. Table 2, based on Kaastra \& Mewe (1993), summarizes the average charge states realized for the M, L, and K shell photoionization of the heavy element atoms and ions considered in this work. As an example, for the (outer) L-shell ionization of C ( X-ray energy below 280 eV), there are no Auger electrons and singly-ionized C$^+$ is the result. However, when an inner K-shell electron is ejected (X-ray energy above 280 eV), an Auger electron is also generated and the result is C$^{++}$. 
 
\begin{center}
Table 2
\nopagebreak

\vspace{-2ex}
Ion Charge States$^\mathrm{a}$
\nopagebreak

\vspace{-1ex}
\begin{tabular}{lccc}    
\hline
\hline
Element 	& M 	& L 	& K 	\\        
\hline
C	& \dots	& 1	& 2	\\
N   & \dots	& 1	& 2	\\
O   & \dots	& 1	& 2	\\
Ne  & \dots	& 1	& 2	\\
Na	& 1		& 2	& 2	\\
Mg  & 1		& 2	& 2	\\
Si  & 1		& 2	& 2	\\
S 	& 1		& 2	& 2	\\	 
Ar	& 1		& 2	& 2	\\
K	& 2		& 2	& 2	\\
Fe	& 2		& 2	& 2	\\
\hline
\end{tabular} \\
$^\mathrm{a}$ Number of electrons produced by direct ionization of neutral atom.
\end{center}
 
We use Table 2 to define the following chemical reactions for the direct ionization of particular atomic shells, which are added to the chemical network: 
\begin{align}
\hspace{2.1in}
\ha  &\xrightarrow{\zeta_{\rm K}} \happ + 2e   \nonumber  \\
\hap &\xrightarrow{\zeta_{\rm K}} \happ + e  \nonumber  \\
\ha &\xrightarrow{\zeta_{\rm L}} \hap + e \hspace{0.5in} \mathrm{A = C,N,O,Ne}  \nonumber \\
\ha &\xrightarrow{\zeta_{\rm L}} \happ + 2e \hspace{0.4in} \mathrm{A \ge Na}  \nonumber  \\
\hap &\xrightarrow{\zeta_{\rm L}} \happ + e   \nonumber \\
\ha &\xrightarrow{\zeta_{\rm M}} \hap + e \hspace{0.5in} \mathrm{A = Na, Mg, Si, S, Ar}  \nonumber 
\\
\ha &\xrightarrow{\zeta_{\rm M}} \happ + 2e \hspace{0.4in} \mathrm{A = K, Fe}  \nonumber  \\
\hap &\xrightarrow{\zeta_{\rm M}} \happ + e 
\end{align}
Doubly-charged ions  $\hatwop$ are produced in all cases except for L-shell photoionization of C, N, O and the M-shell ionization of Ne, Na, Mg, Si, S, Ar. As discussed earlier, charge states greater than two have been consolidated into $ \hatwop$ because of rapid charge exchange with atomic and molecular hydrogen and other species. 

The ionization rates in the reactions above, for the increase in the charge state of A or $\hap$ using the partial rates for each shell, are:
\be
\label{M_ionA}
\zeta_{\rm M} = \int_{E_0}^{E_{\rm L}}\, 
dE\, F(E) \, \sigma_{\ha}(E)\,,
\ee  
\be
\label{L_ionA}
\zeta_{\rm L} = \int_{E_{\rm L}}^{E_{\rm K}}\, 
dE\, F(E) \, \sigma_{\ha}(E)\,,
\ee  
\be
\label{K_ionA}
\zeta_{\rm K} = \int_{E_{\rm K}}^{\infty}\, 
dE\, F(E) \, \sigma_{\ha}(E)\,.
\ee  
The total direct ionization is then the sum
\be
\zeta_{\rm A, \, dir}  = \zeta_{\rm K} + \zeta_{\rm L} + 
\zeta_{\rm M}.
\ee

To summarize this section, we recall that general formulae were given in Sec.~2.1 for the direct and secondary ionization rates for atom A in Eqs.~\ref{zetadirA} and \ref{zetaAsec} as well as formulae for the direct and secondary ionization rates per H nucleus in Eqs.~\ref{tot_dir} and \ref{totzetasec}. In this section we introduced the approximation, $\zeta \simeq  \zeta_{\rm sec}$ and obtained practical expressions for the ionization rate, Eq.~\ref{approxtotzeta}, and the secondary ionization rate for atom A, Eq.~\ref{zetaAsec}. 

\subsection{Ion Reactions \& Chemistry } 

The heavy ion abundances in protoplanetary disk are determined by the X-ray production rates just discussed and by  the chemistry of the ions. The abundances are also determined by the gas-phase elemental abundances, i.e., by the poorly known depletion in protoplanetary disks. We consider the heavy elements with cosmic abundances $ > 10^{-6}$: C, N, O, Ne, Na, Mg, S, Si, Ar, K and Fe, but without calcium and nickel, which are likely to be highly depleted in disk gas as they are in the interstellar medium (Savage \& Sembach 1996; Jenkins 2009).
This set is somewhat larger than the recent disk chemistry model of (Walsh et al.~2010) 
and many previous studies that include Mg, S, Si and Fe (Woitke et al.~2009, Ilgner et al.~2006, Markwick et al.~2002, Le Teuff et al.~2000, Willacy et al.~2000; Willacy \& Langer 2000). We also include the two atoms with the lowest ionization potentials (IP) that can be collisionally ionized (Na an K) and the nobles gases that are important for diagnostic purposes (Ne and Ar).   

The ionization theory presented here is a more detailed version of the classic work of Oppenheimer and Dalgarno (1974; henceforth OD74) on the ionization of dense interstellar clouds. One of their main ideas was that many heavy-metal elements such as Mg, Ca, Na and Fe are largely chemically inert, and their ions have the potential to be relatively abundant where they are destroyed mainly by weak electronic recombination. They introduced the concept of a generic heavy atom to represent the effect of all such atoms, whereas we develop the ionization chemistry for eleven heavy atoms. OD74 did not discuss the ions of relatively abundant neon and argon or any multiply-charged ions, none of which would be likely to be found in the interstellar medium except in the proximity of an  X-ray source,  as in Meijerink \& Spaans (2005). In this paper we treat Ne, Na, Mg, S, Si, Ar, K and Fe in disk gas exposed to X-rays. X-ray ionization is calculated following Sec.~2. Electronic recombination rate coefficients for Si, S, K and Fe are from Landini \& Monsignori-Fossi (1991) and from Badnell (2006) for lighter elements. Reactions of heavy ions with  H and $\hm$ were not discussed by OD74; they can play an important role in protoplanetary disks. 

Heavy-element ions may also react strongly with abundant neutral molecules such as CO and $\htwoo$. Singly-ionized ions may be produced by proton transfer from abundant molecular ions such as $\hthreep$, $\hcop$ and ${\rm H}_{3}{\rm O}^{+}$ to the heavy atoms, which generally have high proton affinities. The review by Anicich (1993) remains a valuable resource for the relatively scarce data on the reactions of singly-ionized heavy atoms. The situation for doubly-ionized ions is more difficult, but there is enough useful information scattered throughout the literature to obtain a reasonable preliminary picture of their reactivity. In the next section we will report on thermal-chemical calculations that illustrate some of the consequences of our theory of X-ray ionization of heavy atoms. Among the approximately 1200 chemical reactions in our program\footnote{Available on request from the lead author.}, about 100 relate to heavy ion reactions for Ne, Na, Mg, S, Si, Ar, K and Fe. A complementary study of the molecular abundances obtained with this program can be found in  Najita,  \'{A}d\'{a}mkovics \& Glassgold (2011). 

Ion reactions with abundant H or $\hm$ can be important, as we demonstrated earlier for the neon ion (Glassgold, Najita \& Najita 2007; GNI07).  Even for a rate coefficient coefficient as small as $10^{-14} \ccps$, the reaction,
\be
\netwop + \h \rightarrow  \nep + \hp
\ee
can dominate the destruction of $\netwop$ at moderate depths in a protoplanetary disk atmosphere. Theoretical calculations suggest that charge transfer of the double ions of Ne, Na, Mg, S, Si, Ar, K and Fe to H is generally weak, except for Si$^{++}$ (Gargaud et al.~1982) and Fe$^{++}$ (Neufeld \& Dalgarno 1987), where the rate coefficients are of order $10^{-9} \ccps$ in the temperature range of interest. Both of these ions are complex and manifest the lowest endothermicity ($\sim$2.5\,eV) for charge exchange of these ions with atomic H. Turning to the interactions with molecular hydrogen, four outcomes are possible:
\begin{align}
\label{doubles}
\hatwop + \hm  \rightarrow \hspace{0.25cm}& \hap    + \hmp      \nonumber \\ 
			  							 & \hap    + \h + \hp  \nonumber \\ 
                          				 & \ha      + 2\hp      \nonumber \\
			 							 & \ha\h^+  + \hp.              
\end{align}
Charge exchange without dissociation is always allowed energetically for the elements under consideration, but all four channels are open in most cases. The exceptions occur for Mg$^{++}$, Si$^{++}$ and Fe$^{++}$, where only the production of $\ha \hp$ and charge exchange without dissociation are possible. The production of $\ha \hp$ leads to rapid loss of ionization by dissociative recombination. Among the heavy double ions of interest, laboratory experiments have only been reported for S$^{++}$  (Chen, Gao \& Kwong 2003)  and C$^{++}$ (Gao \& Kwong 2003), where the rate coefficients are large. We assume that all of the rate coefficients for the reaction of doubly-charged ions and $\hm$  are large, and, in the absence of any information on branching rations, we assume that they are equal.

In our simplified ion-abundance chemistry, we have not included all charge transfer reactions between singly-ionized atoms and neutral atoms with lower IP.  Such reactions are often assumed to be fast, as in the UMIST database (Woodall et al.~2007), even though there are very few relevant experiments or calculations. This assumption is most reasonable when the IP differences are small and the reactants are complex, e.g., for Si$^+$ + Fe and Fe$^+$ + Mg with ionization potential differences $\sim$0.3\,eV, which we test in our reaction network and discuss in Section 3.5.

The general theory presented in Section 2 assumes that the dominant form of every element is atomic, neutral or ionic. However, beyond a certain depth into the disk, typically for vertical column densities $\Nh > 10^{21} \psqcm$, many of the lighter elements are incorporated into molecules, as discussed below in Sec.~4.3. The absorption of an X-ray by a molecule usually leads to dissociative ionization, and this process must be treated in order to calculate atomic ion abundances in molecular regions. For $\hm$, this has already been done in our program, as described in GNI04 (Eq.~10). One obstacle in understanding the effects of dissociative ionization of heavy molecules by X-rays is the lack of experimental information on branching ratios, in contrast to molecular X-ray absorption cross sections which are well approximated as the sum of atomic cross sections. Although high-resolution measurements exist in the immediate neighborhood of the K-edge for selected molecules (e.g., Piancastelli et al.~1999), the branching ratios are not the same at energies away from the edges. Indeed, the resonance-like structure measured in X-ray absorption cross sections close to photoionization thresholds are not important for astrophysical applications because they occur over a small energy range compared to the broad band of X-ray energies emitted by YSOs. 

Unlike the case of atoms, where double ions are the main product of ionization by keV-energy X-rays, the dominant channel for simple molecules is fragmentation into singly-ionized species. For example, Hitchkock's measurements of CO just below 1 keV (well above the C and O K-edges) yield the following rough branching ratios: C$^+$ + O$^+$ (70\%), C + O$^{2+}$ (10\%),  O + C$^{2+}$ (10\%), C$^+$ + O$^{2+}$ (5\%) and O$^+$ + C$^{2+}$ (5\%) (Hitchcock et al.~1988). However in molecular regions, double ions are rapidly destroyed by reaction with $\hm$ and other species, independent of how they are formed. The net effect of X-ray ionization of heavy molecules is the production of singly-charged atomic ions. A reasonable approximation to the direct dissociative ionization of heavy molecules can be obtained by assuming, as it is done in the literature, that the molecular X-ray absorption cross section is the sum of the atomic cross sections,  and that the branching ratios can be all taken to be zero except for the singly-ionized channels. In cases where the branchings have been measured or calculated, as in the case of CO, they can be used, instead of assuming that the outcome is only single-ions. Collisional ionization of molecules by the secondary electrons also occurs and leads primarily to single ions. We have found that X-ray dissociation of CO, where the branchings are reasonably well known, has some quantitative effects on the abundances of several atomic and molecular species, but the overall nature of the thermal-chemical transition is unchanged.

\section{Results} 

\subsection{The Local X-ray Spectrum}

The primary photoionization rate for an atom in Eq.~\ref{zetapriA} depends on the local X-ray flux spectrum, $F(E)$ and the photoionization cross-section. At a particular location $(R,z)$ in the disk, $F(E;R,z)$ is determined by the incident flux in the absence of the disk, $F_0(E;R,z)$, and by the attenuation by the disk material, ignoring scattering. We use cylindrical coordinates, assuming an azimuthally symmetric disk. Satellite observations of YSOs provide us with information on the X-ray luminosity $\LX$ and its spectrum $\LX(E)$, but little about the origin of the X-rays. We assume that they originate close to the young star so that,
\be
\label{stellarX}
F_0(E;R,z) = \frac{\LX(E)}{R^2 + z^2} \hspace{0.5in} E > E_0,
\ee  
where $E_0$ is a low energy cut-off that takes into account the attenuation of the stellar X-rays prior to their encountering the disk, e.g., by the accretion funnels and the wind of the YSO. For purposes of illustration, we represent the stellar X-ray spectrum by a single X-ray temperature, $T_{\mathrm{X}}=1$\,keV, a luminosity, 
$L_{\mathrm{X}}=2\times10^{30} \ergps$, a low-energy cutoff, $E_{0}=$\,200\,eV, and an upper cutoff of 20\,keV.
Ignoring scattering, the local X-ray spectrum is obtained from the source spectrum as
\be
\label{Flocal}
F(E;R,z) = e^{-\tau(E;R,z)} \, F_0(E;R,z) 
\ee
where $\tau(E;R,z)$ is the X-ray opacity along the line of sight through the disk, i.e., the integral of the volume density of hydrogen nuclei $\nh$ times the average cross-section in Eq.~\ref{avxsec}. For the density distribution, we use the D'Alessio et al.~(1999) model of a typical T Tauri star flared accretion disk, with the gas properties determined by X-ray irradiation following GNI04. The parameters of the model are given in Table 3.

\begin{center}
Table 3 
\nopagebreak

\begin{tabular}{lcl}    
\multicolumn{3}{c}{Disk Model Parameters } 			\\
\hline
\hline
Parameter 		& Symbol		& Value					\\        
\hline
Stellar Mass		& $\Mst$		& $0.5\Msun$      			\\
Stellar Radius		& $\Rst$		& $2\Rsun$	       			\\
Stellar Temperature	& $T_*$			& 4000\,K	       			\\
Disk Mass		& $M_D$			& $0.005\Msun$       			\\
Disk Accretion Rate	& $\dot M$		& $10^{-8} \Msun {\rm yr}^{-1}$   	 \\
Dust Size		& $a_g$			& $0.707 \micron$			\\
X-ray Luminosity 	& $\LX$			& $2\times10^{30} {\rm erg} \ps$ 		\\
X-ray Temperature	& $T_{\rm X}$		& 1\,keV		         	\\
Mechanical Heating	& $\alpha_h$		& 1.0	                      	\\
\hline
\end{tabular}
\end{center}
Here $\alpha_h$ is a phenomenological parameter that specifies the strength of the mechanical heating defined in Eq.~(12) in GNI04, and $a_g$ is the geometric mean of the minimum and the maximum grain radii in an MRN dust size distribution; the grain surface area per unit volume is proportional to $1/a_g$ and to the dust-to-gas ratio of the disk.
Our choice for $\LX$ is close to the the median K star X-ray luminosity in the 
Orion Nebular Cluster (Preibisch \& Feigelson 2005).

In order to evaluate $\tau(E,R,z)$ in Eq.~\ref{Flocal}, we use the D'Alessio tables of vertical mass densities as follows: (1) The mass densities are interpolated onto a logarithmic grid of disk radius $R$ and altitude $z$ above the mid-plane, covering a radial range of $0.28-200$\,AU and an altitude range of $0.01-316$\, AU. (2) The mass density along the line-of-sight to each grid point is integrated and converted to a line of sight hydrogen nucleus column number density into the disk, $N_{\mathrm{los}}(R,z)$, by dividing with the mean atomic mass, $1.425 \,m_{\rm H}$, where $m_{\rm H}=1.67\times10^{-24}$\,g, is the mass of hydrogen. (3) The opacity is then calculated as
\be
\tau(E,R,z) = N_{\mathrm{los}}(R,z)  \,\bar{\sigma}(E),
\ee
where $\bar{\sigma}$(E) is the abundance-weighted cross-section of neutral atomic species given in Equation~\ref{avxsec}. 
In contrast to higher ionization states, X-ray absorption cross-sections for first and second ions differ little from those for neutral atoms. As noted in Section 2.2 we ignore Compton scattering, which is appropriate for vertical columns approaching 10$^{24}$\,cm$^{-2}$. Substituting the opacity into Equation~\ref{Flocal} and dropping the location subscripts for clarity gives the local X-ray flux spectrum in terms of the incident X-ray flux and the disk line of sight column densities,
\be
F(E) = e^{- N_{\mathrm{los}} \bar{\sigma}(E)}  F_0(E).
\ee	
Figure 1 presents the X-ray number spectra from $E=13.6$\,eV to 20 keV for a range of vertical column densities into the disk (from above) at a radial distance 1\,AU. The figure can be used to find the typical X-ray energy at a given column density. The EUV photons are of course absorbed out at very small columns. The dashed line gives the position of the lower energy cutoff used in the rest of this paper, $E_0 = 200$\,eV. 
\subsection{Ionization Rates in the Disk} 

Once $F(E)$ is determined, we can calculate the dominant ionization rate $\zeta$ using Equation~\ref{approxtotzeta} in Section 2.3 and the direct ionization rates for individual atoms in Section 2.4. These calculations not only require a disk structural model, but the specification of the poorly constrained heavy-atom elemental abundances.  The only extensive measurements of gas-phase abundances of heavy elements come from UV absorption line spectroscopy of relatively thin interstellar clouds ($A_V \lessapprox 2$\, mag.), and these abundances vary from cloud to cloud. For purposes of illustration, we use the measurements for the well studied cloud $\zeta$ Oph (Savage \& Sembach 1996.~Table 1). They are listed in Table 4. The last column shows depletion factors, defined as the ratio of solar to interstellar abundances, with the former coming from Table 1 of the review of solar abundances by Asplund et al.~(2009). As discussed in this paper and in Jenkin's (2009) review of interstellar abundances, both sets of abundances have significant uncertainties, in some cases factors of a few. Jenkins highlights the case of S, where the UV observations suggest essentially no depletion in interstellar clouds, despite the fact that this element is seen in meteorites.

\begin{center}
Table 4 \\
\begin{tabular}{ccc}    
\multicolumn{3}{c}{Elemental Abundances} \\
\hline
\hline
Element 	& Abundance		& Depletion   \\        
\hline
H	& 1.00		& 1.0	\\
He	& 0.10		& 1.0	\\
C	& $1.40 \times 10^{-4}$	& 2.0	\\
N	& $6.00 \times 10^{-5}$	& 1.1	\\
O	& $3.50 \times 10^{-4}$	& 1.4	\\
Ne	& $6.90 \times 10^{-5}$	& 1.2	\\
Na	& $2.31 \times 10^{-7}$	& 7.5	\\
Mg	& $1.00 \times 10^{-6}$	& 40		\\
K	& $8.57 \times 10^{-9}$	& 12		\\
Si	& $1.68 \times 10^{-6}$	& 19		\\
S	& $1.40 \times 10^{-5}$  & 1.0	\\
Ar	& $1.51 \times 10^{-6}$	& 1.7	\\
Fe	& $1.75 \times 10^{-7}$	& 180	\\
\hline
\end{tabular}
\end{center}

We use the K, L, and M-shell energies given by Verner \& Yakovlev (1995). The ionization rates for individual species are plotted in Figure 2 at $R=$1\,AU. All of the rates decrease with increasing vertical column due to attenuation of the X-rays (Figure 1), but the relative importance of K-shell ionization is affected by the increase with atomic number of the location of the K shell edge. On the other hand, K-shell ionization always dominates at sufficiently large column densities because the X-ray spectrum hardens with increasing column density, again as shown in Figure 1. The rates in Figure 2 are roughly 2.5 dex higher than GNI97 in the optically thin limit. This is due to our use of a larger upper-energy cutoff ($20$\,keV), a smaller lower-energy cutoff (200\,eV), and a factor 20 increase in the X-ray luminosity to $\LX= 2\times10^{30}$. The choice of lower-energy cutoff is affected by the nature of the X-ray source and the amount and distribution of close-in circumstellar material. GNI97 used a 1\,keV cutoff, and this high value for the lower-cutoff caused a large underestimate of the unattenuated ionization rate, whereas GNI04 and Glassgold et al.~(2009) used $E_0=$100 and 200\,eV, respectively. In applications we continue to use $E_0=$200\,eV. Photons of this energy will be attenuated by line-of-sight columns of 10$^{20}$ -- 10$^{21}$\,cm$^{-2}$.  Curves for $R^2 \zeta$ are plotted in Figure 3 as a function of vertical column $N_{H}$ for radial distances $R$ between 0.25 and 20\,AU. The radial dependence of $\zeta$ is close to $1/R^2$ until the disk opacity becomes significant, i.e., at vertical columns $\sim10^{21}\,\psqcm$. At larger vertical columns the ionization rate decreases due to attenuation by disk material along the line of sight 
and ionization rates are less than predicted by $1/R^2$ dilution alone.

\subsection{The Thermal-Chemical Transition} 

The gas and dust temperatures are plotted against vertical column density in Figure 4. The dust temperature, $\Td$, is an input for our model from D'Alessio et al.~(1999). It decreases with both radius and vertical column density. At 0.25\,AU  $\Td > $ 800\,K at the top of the atmosphere, dropping to $\sim$300\,K for $\Nh = 10^{24} \psqcm$. Further out in the disk at $R=$20\,AU, the maximum $\Td$ is just over 100\,K, and it drops to below 40\,K at large columns. At each radius the overall trend of $\Td$ with $\Nh$ is similar, with each curve offset to a lower temperature as $R$ increases.

The gas temperature, $\Tg$, is calculated by the thermal-chemical program described in GNI04. The heating agents are X-rays and accretion, and the cooling is mainly by atomic and molecular lines. An essential ingredient is the collisional coupling of the gas and the dust, which in our model generally results in gas cooling. Like the dust temperature, the gas temperature decreases with both radius and downward vertical column. It is almost always larger than the dust temperature, except at large column densities where the gas and the dust are tightly coupled thermally and then the two become equal. In the upper atmosphere, $\Tg$ is the range $\sim$4000 - 6000\,K. It undergoes a steep drop at vertical columns near $10^{21} \persqcm$ to within a factor of two, due to a thermal-chemical transition that involves greatly increased molecular abundances. This transition is not only marked by a sharp decrease in temperature, but by jumps in the abundances of major molecules like $\hm$, CO and H$_2$O, and by a decrease in the electron fraction $\xel$ (as seen Figure 4). All of these changes are mutually supportive, e.g., molecule formation increases the cooling and helps decrease the temperature, while the lower temperature aids molecular synthesis by suppressing reactions that destroy molecules. The overall result is that the transition shown in Figure 4 divides the disk into three regimes: a hot atomic layer on top ($\Nh < 10^{21}$), a warm atomic to molecular transition region ($\Nh \sim 10^{21}- 10^{22}$), and a cool disk region at the bottom ($\Nh > 10^{22}$). Analogous transitions are found in models of shocks (e.g., Hollenbach \& McKee 1989) and photodissociation regions (e.g., Hollenback \& Tielens 1999) in the interstellar medium.

In the hot atmosphere of the disk, the dominant ions are $\hp$ and $\hep$. The variation in electron fraction can be simply approximated by balancing the dominant production via X-ray ionization and destruction by radiative recombination, ignoring $\hep$,
\be
n(\rm H)\, \zeta = \nhp \,\nel \, \alpha
\ee 
where $\alpha$ is the radiative recombination rate. Converting from densities to fractional abundances, using $\xhp \approx \xel$ and $\xh \approx 1$ and rearranging gives an expression for the electron fraction
\be
\xel = \sqrt{\frac{\zeta}{\alpha \, n}},
\ee
familiar from the theory of cosmic ray ionization of the interstellar medium (Spitzer 1978). The electron fraction is determined by the square root of the ionization parameter, $\zeta/n$. The model calculations of $\xel$ are plotted vs the ionization parameter in Figure 5. In the hot atmosphere of the disk, $\xel$ follows the square-root behavior up to a vertical column $10^{21} \persqcm$. Above the transition, the effective recombination coefficient is close to that for radiative recombination of $\hp$, whereas below it is somewhat larger than the recombination coefficients of singly-charged heavy atomic ions. This excess is a reflection of the role of molecular ions in reducing $\xel$ by fast dissociative recombination.

At the transition, there is a steep drop in $\xel$ due to the formation of molecules and molecular ions. The latter are especially important, even though they are less abundant than the atomic ions by at least one order of magnitude, because the rate coefficient for dissociative recombination of molecular ions is typically $10^5$ larger than the rate coefficient for radiative recombination of atomic ions. There are many molecular ions in our chemical model. Which is most abundant depends somewhat on location, e.g., in the warm region they are $\hthreep$, $\HHHOp$, $\hcop$ and N$_2$H$^+$, whereas at $\Nh = 10^{24} \psqcm$ metal hydride ions such as MgH$^+$ and SiH$^+$ also compete in destroying electrons. According to Figure 5, the main thermal-chemical transition occurs at fixed $\zeta/n$. Going deeper, $\xel$ at first varies roughly as the cube root of $\zeta/n$ and then recovers the familiar square-root dependence for $\Nh > 10^{22}$. The extreme sharpness of the transition is due to the major role of mechanical heating associated with the MRI in the GNI04 model. According to Eq. 12 of that paper, the volumetric heating rate is proportional to  $\alpha_h \rho c^2$, where $\alpha_h$ is a phenomenological constant (chosen here to be unity), $\rho$ is the mass density, and $c$ is the sound speed. Since $c^2 \propto \Tg$, there is a potential for runaway since decreasing $\Tg$ reduces the heating which leads to a further drop in $\Tg$. When the viscous heating parameter $\alpha_h$ is decreased, the feedback is weakened, and calculations show that the transition occurs more smoothly. 
         
\subsection{Charged Species in the Disk}   

The ionization state populations of all of the heavy elements included in this work 
are plotted in Figure 6 
vs.~$\Nh$ for $R=1$\,AU. The chemically active species are shown in the top panels, 
the rare gas and alkali earth elements in the middle panels, and the less chemically 
active low-IP elements at the bottom. Neutral and singly-ionized ions are found in 
the hot upper atmosphere, with the neutral state roughly dominating for the high-IP elements (i.e., C, N, O, Ne, Ar) and the singly-charged state for low-IP elements 
(i.e., Na, Mg, Si, Fe). The abundances of both singly- and doubly-charged ions drop 
sharply at the thermal-chemical transition,  with the exception of the low-IP elements, where the singly-charged ions decline smoothly with increasing vertical column. The 
sharp drop in ion abundances at the transition is due to the enhanced chemical activity 
of the high-IP elements, including neon and argon; Ar$^+$ is rapidly destroyed by $\hm$, and Ne$^+$ reacts strongly with H$_2$O. The sulfur missing below the transition is in SO$_2$, and the missing nitrogen is in N$_2$. The molecular aspects of this model are discussed in a companion paper by Najita et al.~(2011). The abundance of all of the 
doubly-charged ions declines dramatically at the molecular transition because they are rapidly destroyed by reactions with abundant $\hm$. Thus significant amounts of doubly-charged ions are only found in the hot upper atmosphere, and here they are led by 
Ne$^{++}$ and Ar$^{++}$. The abundances of Ar$^+$ and Ar$^{++}$ are the same order 
of magnitude, and they both drop significantly in the  molecular regime.    

The large abundance of some of the neutral atoms in Figure 6 is noteworthy. This is readily understood for the high-IP elements with IP(A)\,$>$\,IP(H) = 13.6\,eV. But the 
results for C, S and K, for which $\xha \geqq \xhap$, may be because our illustrative 
model does not include stellar far ultraviolet (FUV) radiation. Bergin et al.~(2003) showed that the stellar FUV flux of the representative T Tauri star BP Tau at a 
distance of 100\,AU is $\sim$540 times the Habing interstellar radiation field, 
or $\sim$5$\times 10^6$ larger at 1\,AU. When we use this field to estimate the UV photoionization rates for the seven heavy atoms with IP(A)\,$<$\,13.6\,eV, they range 
over two orders of magnitude with a typical value of $10^{-3}\ps$. If we compare 
these estimates with the rates at which the neutral atoms are destroyed by particle reactions (usually by X-ray generated ions) just above the transition at a vertical 
column of $5 \times 10^{20} \persqcm$, the unshielded FUV rates are competitive. 

The positive ions that primarily balance the electrons are plotted in Figure 7. H$^+$ is the dominant species in the hot disk atmosphere, followed by He$^+$. At the transition where $\hm$, CO and H$_2$O are formed efficiently, $\xhp$ and $\xhep$ fall off sharply. The abundant heavy atoms  C, N, O, S and their ions are also incorporated into molecules, leaving the relatively unreactive Si$^+$ as the most abundant ion in the warm transition region. The main silicon ions below the molecular transition are Si$^+$ and SiH$^+$, both produced by molecular ion reactions with neutral Si. Beyond $\Nh \sim 6 \times 10^{23}$, Na$^+$ becomes the most abundant ion deep in the disk. It is formed from neutral Na by charge transfer from a variety of molecular ions. The dominance of Si$^+$ is a specific result of using the depleted interstellar abundances in Table 4 of Savage \& Sembach (1996).

The left panel of Figure 8 summarizes our model results on the distribution of ionization at $R=1$\,AU with the abundance plots of the three types of positive ions, along with the electron fraction. Light atomic ions dominate above the thermal-chemical transition and heavy atomic ions below. The abundances of heavy atomic ions depends of course on the assumed elemental abundances. Many disk chemistry and ionization models employ much lower values than employed here and given in Table 4. For example, the thermal-chemical calculation of Walsh, Millar \& Nomura (2010) use the low-metal abundances from the UMIST database publication (Woodall et al.~2006; Table 8). They are $\sim$100 lower than ours for Na and Fe and $\sim 1000$ lower for Mg, S, and Si. For comparison, the right panel of Figure 8 shows the main classes of ions for the extremely low metal abundances in Woodall et al. The general trend in these two panels are similar. The thermal-chemical transition occurs  in much the same way and at the same depth, although it is slightly sharper in the low-metal case. The main difference is that the controlling ions in the warm region just below the thermal-chemical transition are molecular rather than atomic, with first H$_3$O$^+$ and then HCO$^+$ being the dominant ion. Deeper down, Na$^+$ again is the most abundant ion despite the fact that the total abundance of sodium is $\sim 100$ less. That the electron fractions at the highest columns in Figure 8, $10^{24} \persqcm$, are about the same, is accidental; reducing the heavy element abundances further will lead to smaller electron fractions. At this depth in the atmosphere, the heavy element ions are produced by charge exchange with molecular ions and destroyed by radiative recombination. These conclusions must still be regarded as tentative for the largest column densities because X-ray scattering has been ignored. The chemical effects of small grains and large molecules such as PAHs have also been ignored. Large changes in heavy-element abundances can also affect the chemistry of the warm region most accessible to observations, i.e., the layer that extends from $10^{21} \persqcm$ to $10^{22} \persqcm$. For example, the increased depletion of heavy elements illustrated in Figure 8 leads to an order of magnitude decrease in electron fraction and factors of a few decrease in the abundances of warm columns of CO and H$_2$O. 

\subsection{Charge Exchange \& Collisional Ionization}

The relative importance of the heavy ions is also affected by the chemistry. One interesting possibility is charge exchange with other atoms. Reactions of this type are included in the UMIST database although there is almost no experimental or theoretical information on them at low ion energies ($\lesssim 1$\,eV). We consider the following two heavy atom charge exchange reactions as the most likely because of the small difference (0.25\,eV) between the IP of Si and Fe and of Fe and Mg and the complexity of the iron electronic structure,
\be
{\rm Si}^+ + {\rm Fe} \, \ra \, {\rm Fe}^+ + {\rm Si}, 
\ee  
\be
{\rm Fe}^+ + {\rm Mg} \, \ra \, {\rm Mg}^+ + {\rm Fe}.
\ee
We have incorporated these reactions into our program with large rate coefficients ($\sim 10^{-9} \ccps$). The main effect is that the dominant ion in the warm region is no longer Si$^+$ but Mg$^+$. The overall behavior of the molecular transition and the ionization fraction are unchanged.
Among the many heavy-ion reactions in our program, electronic collisional ionization has been included for the low-IP atoms K, Na, S and Si. The only atoms for which this process is important are Na and K, and much more so for K with it's small IP (4.34 eV). Referring to Figure 6, we see that the abundance of K$^+$ is a little larger than that of K in the hot atmosphere (by $\sim 20-30 \% $). This tendency towards neutrality is a consequence of the large electronic recombination coefficient for neutral K (Landini \& Monsignori-Fossi 1991). In the steady state balance with collisional ionization, 
$\alpha \nel n({\rm K}^+) = k_{\rm coll\,ion} \nel  n({\rm K})$, the electron density cancels out, and the rate coefficients do not change much because the temperature in the hot atmosphere doesn't; thus the K$^+$/K abundance ratio is almost constant. Below the transition, collisional ionization is unimportant because the temperature is too low. In its place, K$^+$ is produced by charge transfer from molecular ions to K and by dissociative charge exchange of K$^{++}$ with $\hm$. The balance with electronic recombination now leads to a larger excess of K$^+$ in the warm region and then to the progressive domination of neutral K with increasing vertical column.

Collisional ionization of low-IP atoms was considered long ago by Pneuman \& Mitchell (1965) in the context of the evolution of the magnetic field in a shielded primitive solar nebula \footnote{The rate coefficients for the collisional ionization of the alkali elements used in this paper grossly underestimate their values relative to current understanding.}. Umebayashi (1983) and Umebayashi \& Nakano (1988) considered the effects of thermal ionization in, respectively, very dense interstellar clouds and the solar nebula, using chemical equilibrium calculations. Gammie (1996) used this idea to determine the inner radius of the dead zone in the MRI to be $\sim 0.1$\,AU (Figure 1 of that paper). The idea that thermal ionization of low-IP atoms makes the MRI active at small radii is now widely accepted in the MRI literature, almost always on the basis of chemical equilibrium. The results shown in Figure 6 do not conform to Saha chemical equilibrium because the disk atmosphere is exposed to strong external radiation and because the pressures are quite low. For example, the pressure at the thermal-chemical transition at $R=1$\,AU is $\sim 1$\,nano-bar. Although the Saha equation gives the correct qualitative dependence of the K$^+$/K abundance ratio on $T$ and $\nel$ (Spitzer 1979, Eq. (2-32)),
\be
\label{Saha}
\frac{x({\rm K}^+)}{x({\rm K})} = \frac{2\times 10^{15}\pcc}{\nel}
                           T^{3/2}\exp(-50,375/T),
\ee
it is quantitatively inaccurate. For example it would predict that overwhelmingly all potassium is in K$^+$ above the transition and in neutral K below, whereas we find K and K$^+$ to be the same order of magnitude, a result that holds throughout the inner disk. Even at the midplane for $R=0.1$\,AU, the pressure for the D'Alessio at al.~(1999) generic T Tauri disk model is less than 1\,mbar. Of course the pressures might well be higher at an early stage of disk evolution with large accretion rates. For revealed disks, however, the ionization state needs to be calculated with the the kind of chemical-ionization theory used in this paper, including collisional ionization for low-IP atoms like K and Na.

To conclude this section, our calculations shed light on the relative importance of electron collisional ionization and X-ray ionization for the low-IP heavy elements. The former dominate in the hot region above the thermal-chemical transition, and the latter below. Deep down in the inner regions of a model T Tauri disk, collisional ionization cannot reverse the tendency towards neutrality for the heavy elements, including K and Na, and the rate of decline in ionization is controlled by X-ray irradiation. The pressures in protoplanetary disks are generally too low for Saha chemical equilibrium to apply.

\section{Discussion and Conclusions} 

We have developed a theory for the generation of single and doubly charged ions in relatively cool gas ($k_{\rm B}T < 1$\,eV) irradiated by keV X-rays, treating the atomic and molecular physics of the recombination of high-energy, multiply-charged ions produced by the Auger effect. The population of the several ionic stages of the heavy atoms is important in determining the overall ionization state of the gas and thus its thermal and chemical properties. We have used the D'Alessio model of a T Tauri star disk to illustrate some of the implications of this theory. The gaseous atmosphere of the model spans a wide range of physical properties. In the inner regions at small radial distances, there is a hot (several thousand K) layer on top of the main part of the disk, which is cool (several hundred K). There is also a warm transition (200-2000\,K) layer between the hot and cool regions where the basic chemical character of the disk changes from atomic to molecular. This is where most of the hydrogen, carbon and oxygen are transformed into the molecules $\hm$, CO and H$_2$O, and where numerous complex molecules are also formed. In the text, we referred to this change as a {\it thermal-chemical transition}. A basic finding of this paper is that the abundances of the doubly-charged ions drop sharply at this transition due to interactions with molecules.  

An immediate deduction from this result is that the spectral lines of some doubly-ionized atomic ions are potentially characteristic diagnostic of the hot layer of protoplanetary disks. On the other hand, the most abundant ions in this region are the light ions, $\hp$ and $\hep$. Singly-charged heavy atomic ions dominate the ionization balance below the thermal-chemical transition, at least for the interstellar depletions used here for purposes of illustration. In this case, Si$^+$ is dominant in the warm transition region, and eventually Na$^+$ takes over this role deeper down in the cool part of the disk. These specific conclusions are, however, sensitive to the assumed abundances of the heavy elements, which are largely unconstrained by observations. If there are fewer heavy elements than we have used, then molecular ions also play an important and frequently dominant role. 

The heavy atomic atoms and ions carry many optical and IR lines that have potential for measuring the heavy element abundances of protoplanetary disks, or at least their ratios. Significant theoretical efforts have already begun in this direction (e.g., Gorti \& Hollenbach, 2004, 2008, 2009; GNI07; Meijerink, Glassgold \& Najita 2008; Woitke, Kamp \& Thi 2009; Kamp et al.~2010; Aresu et al.~2011). In addition to the familiar C\,I 609, C\,II 158 and O\,I 63 and 145 $\mu$m lines, of special interest for {\it Herschel}, and to the Ne\,II 12.8 and Ne\,III 15.5 $\mu$m lines detected by {\it Spitzer}, Gorti \& Hollenbach (2008, 2009) have called attention to the mid-infrared lines of both singly- and doubly-ionized S, Si and Fe. Extensive predictions have been made of optical and infrared line fluxes by Ercolano et al.~(2008, 2009) from the hot disk atmosphere and by Ercolano \& Owen (2010) for photo-evaporative winds, both irradiated by X-rays.  Although outside the scope of this paper, the theory presented here impacts line flux calculations because it affects the abundances of the ions that carry the lines. A concrete example is the Ne\,II 12.8 $\mu$m line, now detected in almost 60 YSOs (Guedel et al.~2010). Theoretical modeling of its emission from protoplanetary disks has been based on a simplified closed-form theory (Glassgold, Najita \& Igea 2007) developed for a purely atomic region. The present theory is more general in several respects, most notably in it's fuller treatment of neon ion interactions with molecules and other species. This can be seen Figure 6, where the abundances of the singly- and doubly-charged ions of both neon and argon decrease abruptly at the transition due to molecular reactions. The relatively high column density warm layer just below the transition no longer contributes significantly to fluxes of lines such as Ne\,II 12.8 $\mu$m, in this case because Ne$^+$ is efficiently destroyed by H$_2$O in the warm layer. Thus, the relatively abundant singly- and doubly-charged ions of neon and argon are diagnostic of the hot layer just above the transition region. The Ne\,II and Ne\,III fine-structure lines have already been detected, and Gorti \& Hollenbach (2009) have called attention to the possibilities of the Ar\,II 6.985 and the Ar\,III 8.991 $\mu$m lines, although the former is difficult to observe due to telluric contamination. In addition to their infrared lines, Ne\,III and Ar\,III have forbidden optical transitions analogous to those of iso-electronic O\,I: $\lambda$3967, $\lambda$3868 and $\lambda$3342 for Ne\,III and $\lambda$7751, $\lambda$7135 and $\lambda$5191 for A\,III. Ar\,III is of particular interest because, according to Figure 6, Ar$^{++}$/Ar$^+$ ratio is of order unity. We plan to return to the calculation of line of atomic lines fluxes based on the results of this paper.

The theory presented here is also relevant to the longstanding problem of the ionization of the dense, largely-shielded regions of protoplanetary disks and to the properties of the MRI dead zone predicted by Gammie (1996). The present work applies only to a small fraction by mass of the active layers of the MRI above and below the dead zone, i.e., to vertical columns densities $< 10^{24} \psqcm$. Already near this depth, Compton scattering is important for extending the range of X-ray ionization, as shown by Igea \& Glassgold (1999). Scattering is not yet included in the program used here for purpose of illustration, nor have other important chemical effects such as reactions with grains and PAHs. In addition to far UV radiation, other ionization sources may also be present (e.g., Turner \& Drake 2009). We plan to extend our theory to apply deeper through the active layers and into the dead zone. 

In conclusion, we wish to acknowledge the inspiration of the pioneering work by Oppenheimer \& Dalgarno (1974), who first tackled the problem of the ionization of dense gas. This work is not only an elaboration of the basic processes that they considered, but it supports their idea of a generic heavy atom that dominates the ionization balance. In the case of a protoplanetary disk irradiated by X-rays, we find that at most locations below the thermal-chemical transition, a single atomic ion does dominate. The identity of the dominant ion changes in a way that reflects the assumed elemental abundances of the heavy elements and the nature of the individual chemical activity induced by the external radiation and heating.  

\acknowledgments

We thank Eugene Chiang for helpful comments before submission as well as the thoughtful suggestions of an anonymous reviewer that have improved the manuscript. This work has been supported in part by NASA grant NNG06GF88G (Origins) and by NASA {\it Herschel} contracts 132594 (Theoretical Research) and 1367693 (DIGIT) to  
UC Berkeley.

\begin{figure} \begin{center}
\includegraphics[width=5.in]{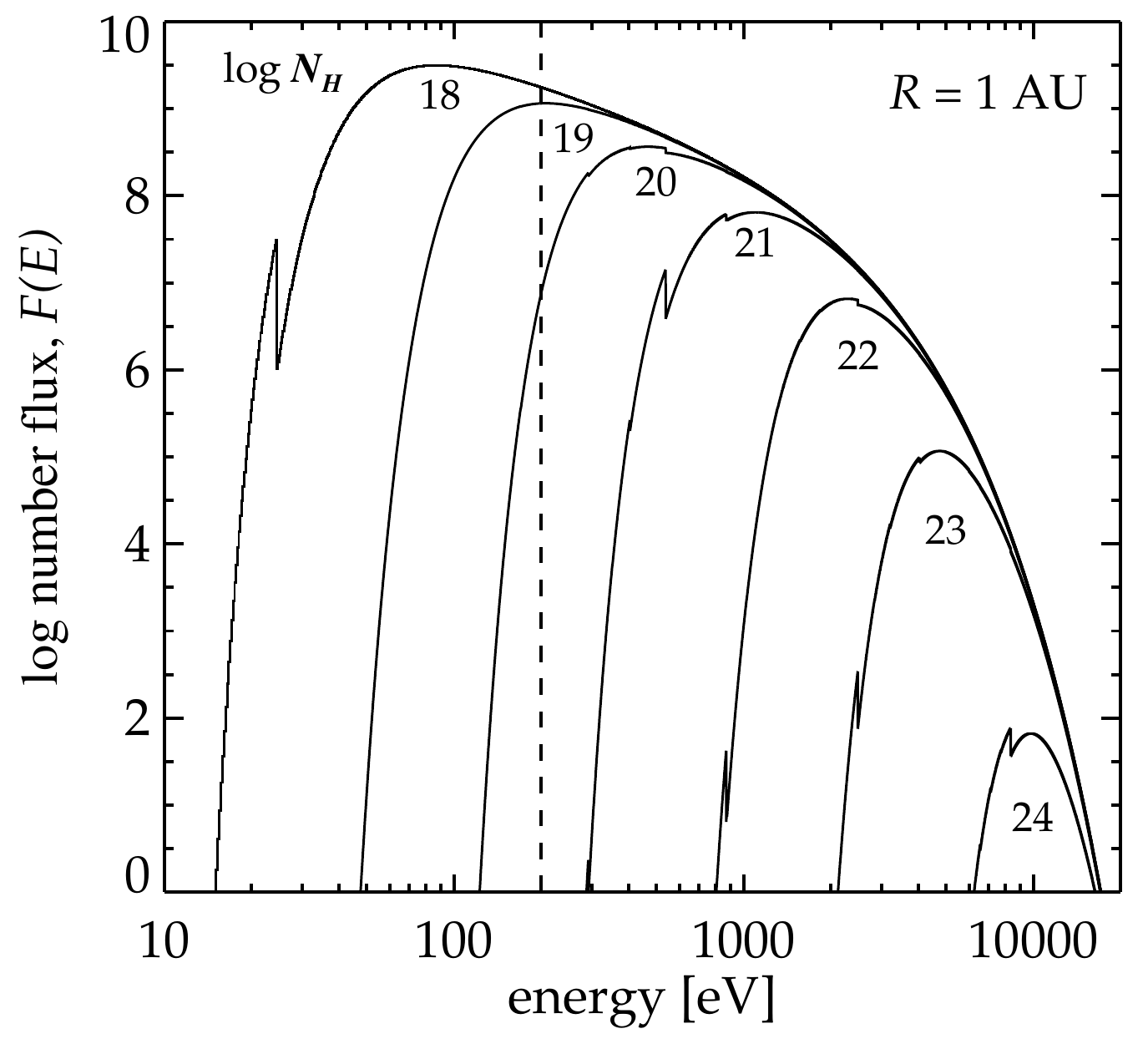}
\caption{The local x-ray number flux spectrum, $F(E)$,  at a radial distance of 1 AU for various altitudes 
above the disk mid-plane. The vertical coordinate is the column density perpendicular to the 
mid-plane, $\Nh$, with curves shown at $\Nh = 10^{18}-10^{24}\,\psqcm$, corresponding to a 
physical altitude of 0.23--0.07 AU. The optical depth is calculated along the line-of-sight from the 
central source.}
\end{center} \end{figure}

\pagebreak

 \begin{figure}
 \includegraphics[width=6.75in]{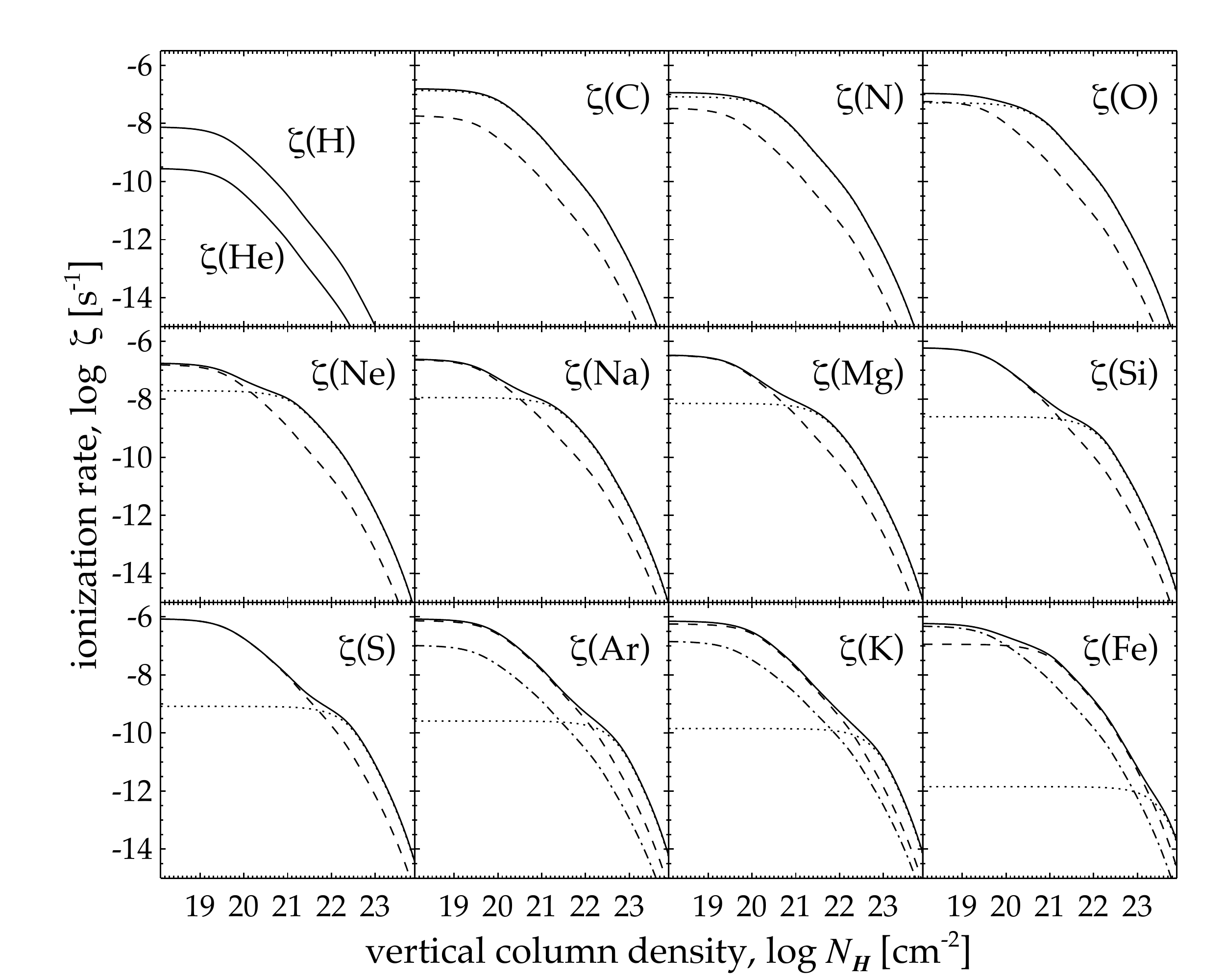}
\caption{Ionization rates per atom, $\zeta(\ha)$, are plotted for each element as a function of 
vertical column density, $\Nh$, at a radial distance of 1\,AU in a typical flared T Tauri disk 
(described in Section 4). The rates for ionization of electrons from specific shells; K (dotted), L 
(dashed), and M (dash-dot), are summed to calculate the total direct rate (solid) for each atom 
according to Eq.~21. }
\end{figure}

\pagebreak

\begin{figure} \begin{center}
\includegraphics[width=5.in]{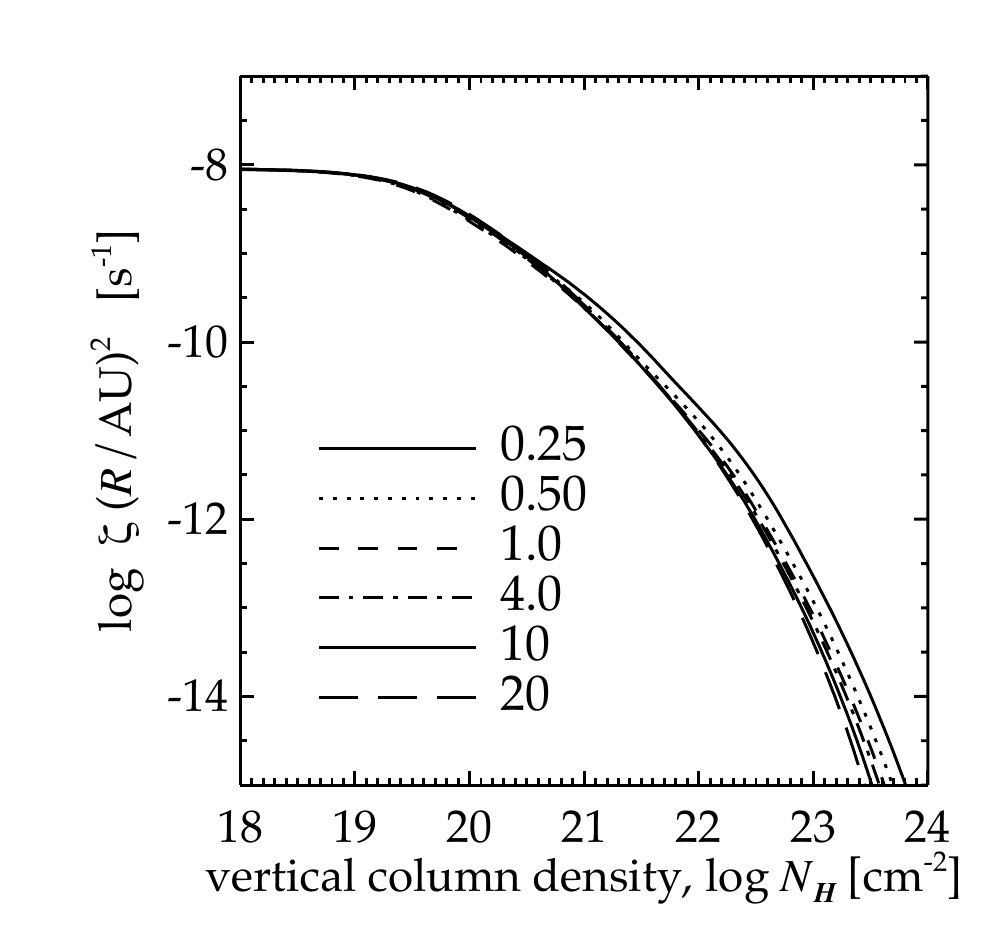}
\caption{Ionization rates, $\zeta$, at various radii, $R$, plotted against vertical column density, $\Nh$. Line styles identify radii in AU as noted in the legend.}
\end{center} \end{figure}

\pagebreak

\begin{figure} \begin{center}
\includegraphics[width=4.in]{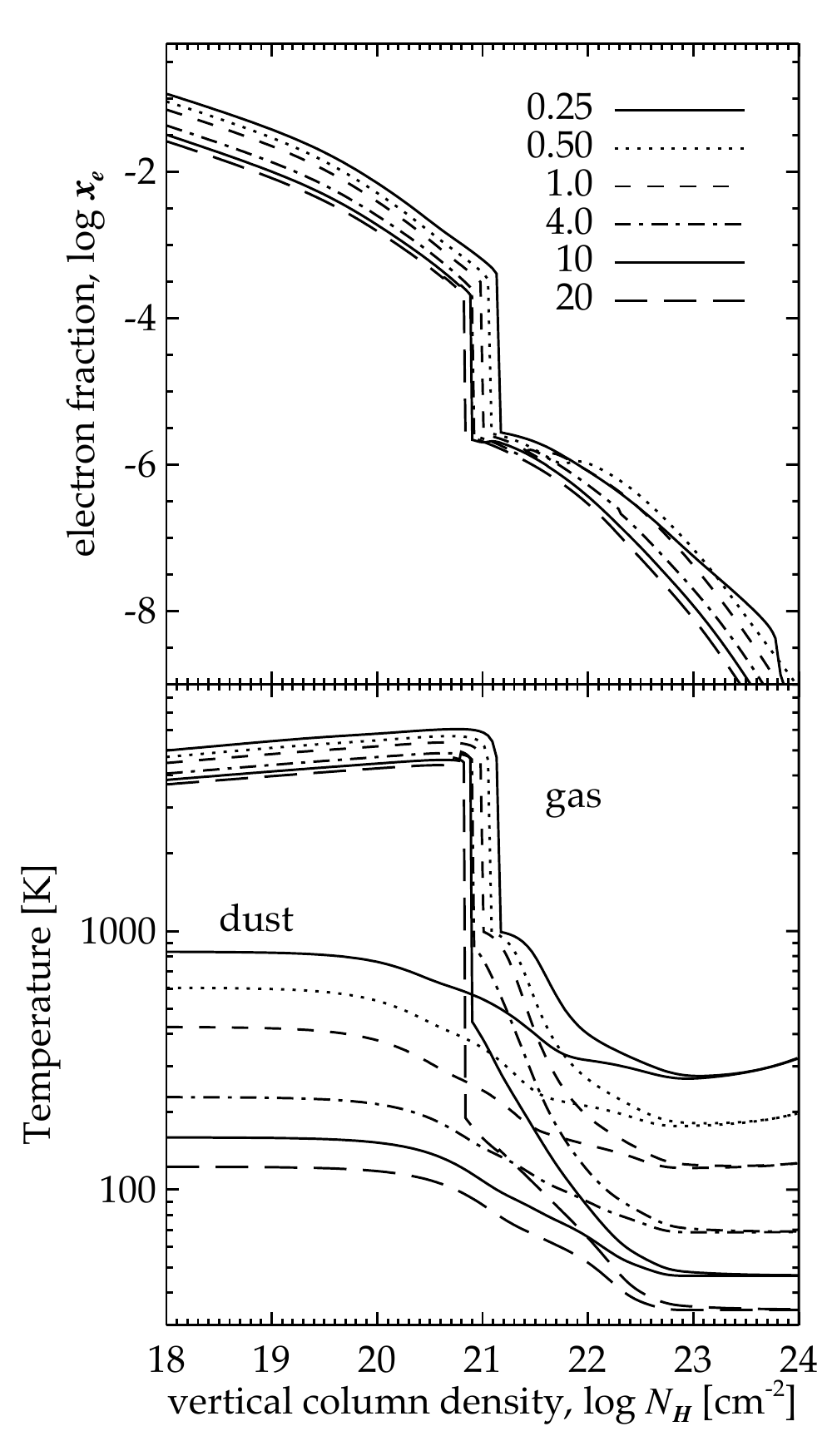}
\caption{Physical properties of the disk atmosphere plotted against vertical 
column density, $\Nh$, at various radii; electron fraction, $\xel$ (top), and the temperatures of gas and dust (bottom).  Line styles identify radii in AU as noted in the legend. }
\end{center} \end{figure}

\pagebreak

\begin{figure} \begin{center}
\includegraphics[width=5.in]{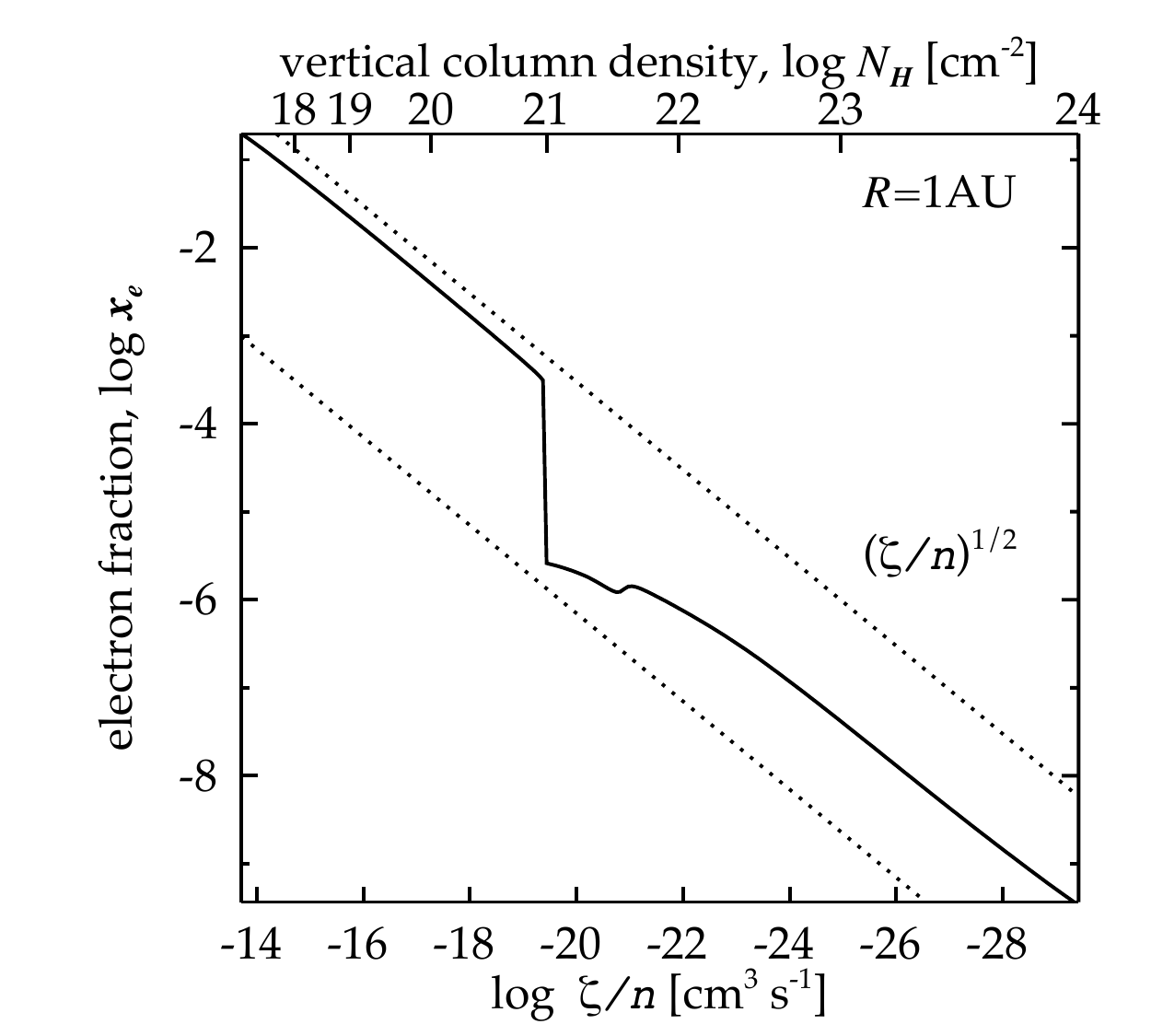}
\caption{The electron fraction, $\xel$ (solid line), is plotted against the ionization parameter $\zeta/n$ 
at 1\,AU.  The dotted lines show an inverse-square dependence for comparison. Except in the 
molecular transition region near log $\Nh=$ 21 to 22, $\xel$ generally varies roughly as $
(\zeta/n)^{1/2}$.}
\end{center} \end{figure}

\pagebreak

\begin{figure}\begin{center}
\includegraphics[width=6.75in]{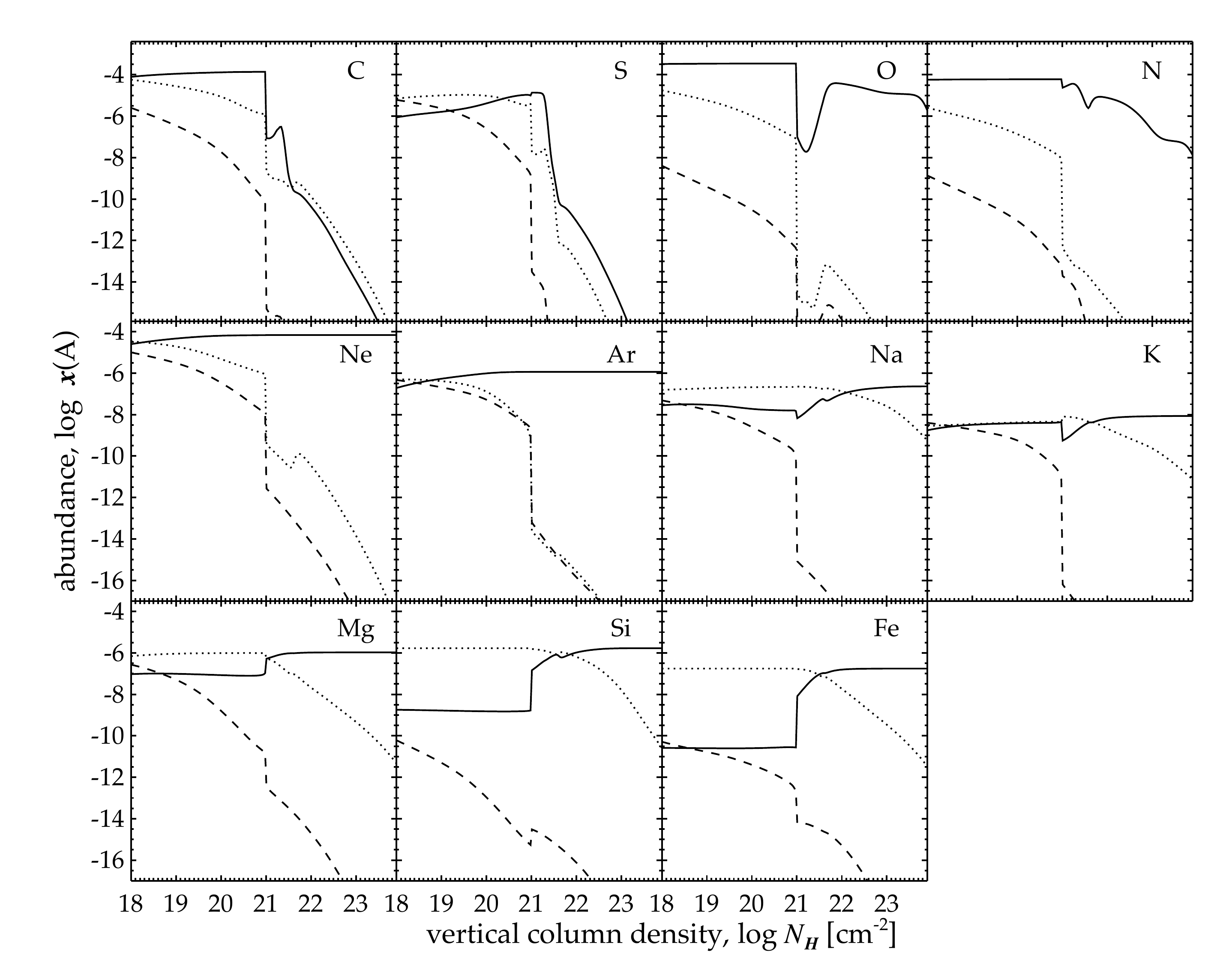}
\caption{Ionization states of the heavy atoms plotted vs.~vertical column density, $\Nh$,  at 1\,AU. The 
neutrals, A,  are represented by solid lines, singly-charged ions, A$^+$, by dotted lines,  and doubly-charged ions, A$^{++}$, by dashed lines.}
\end{center} \end{figure}

\pagebreak

\begin{figure}\begin{center}
\includegraphics[width=5.in]{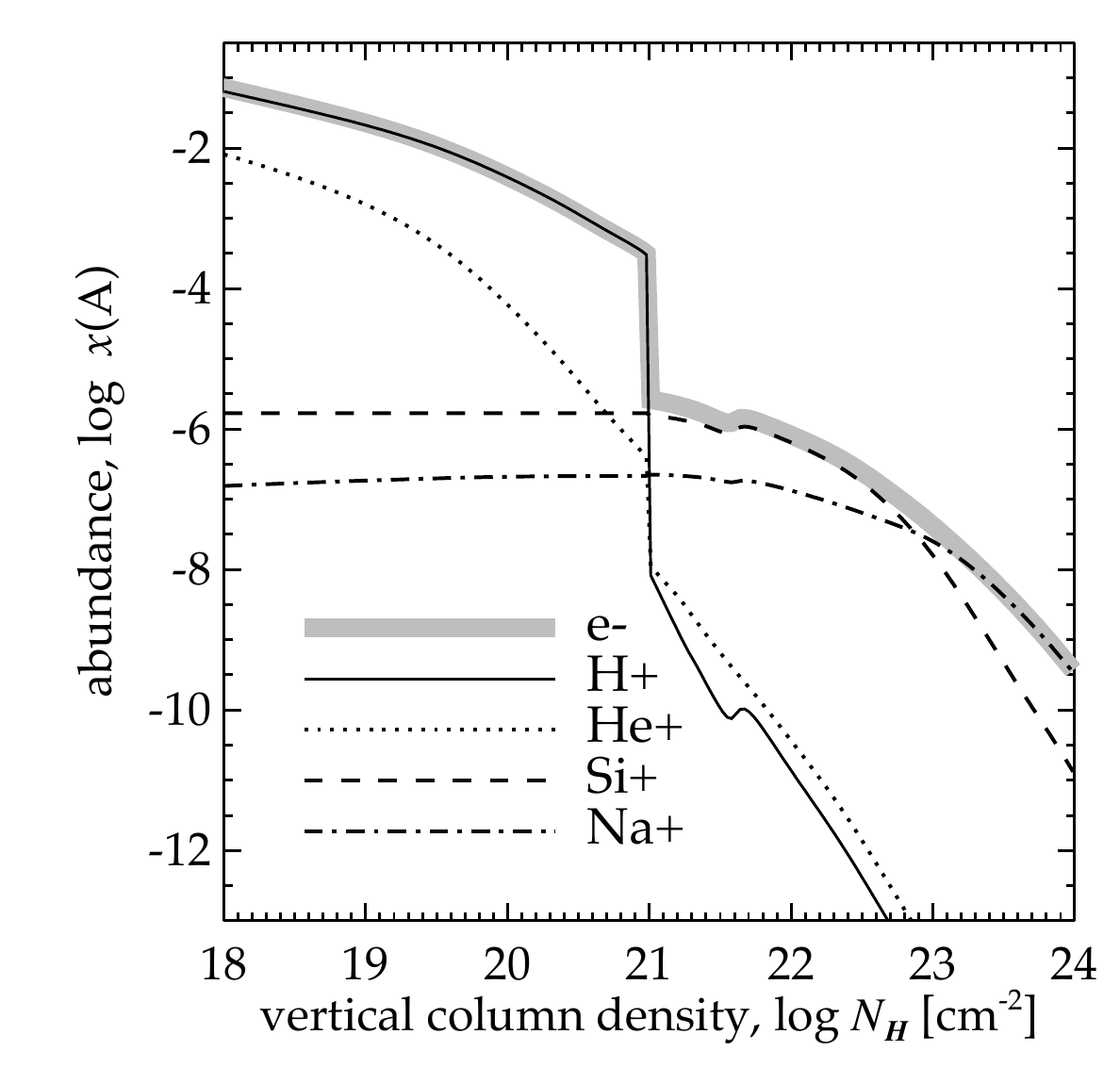}
\caption{The dominant ionic species (identified by line style in legend) plotted vs.~vertical column density at 1\,AU.}

\end{center} \end{figure}

\pagebreak

\begin{figure}\begin{center}
\includegraphics[width=6.5in]{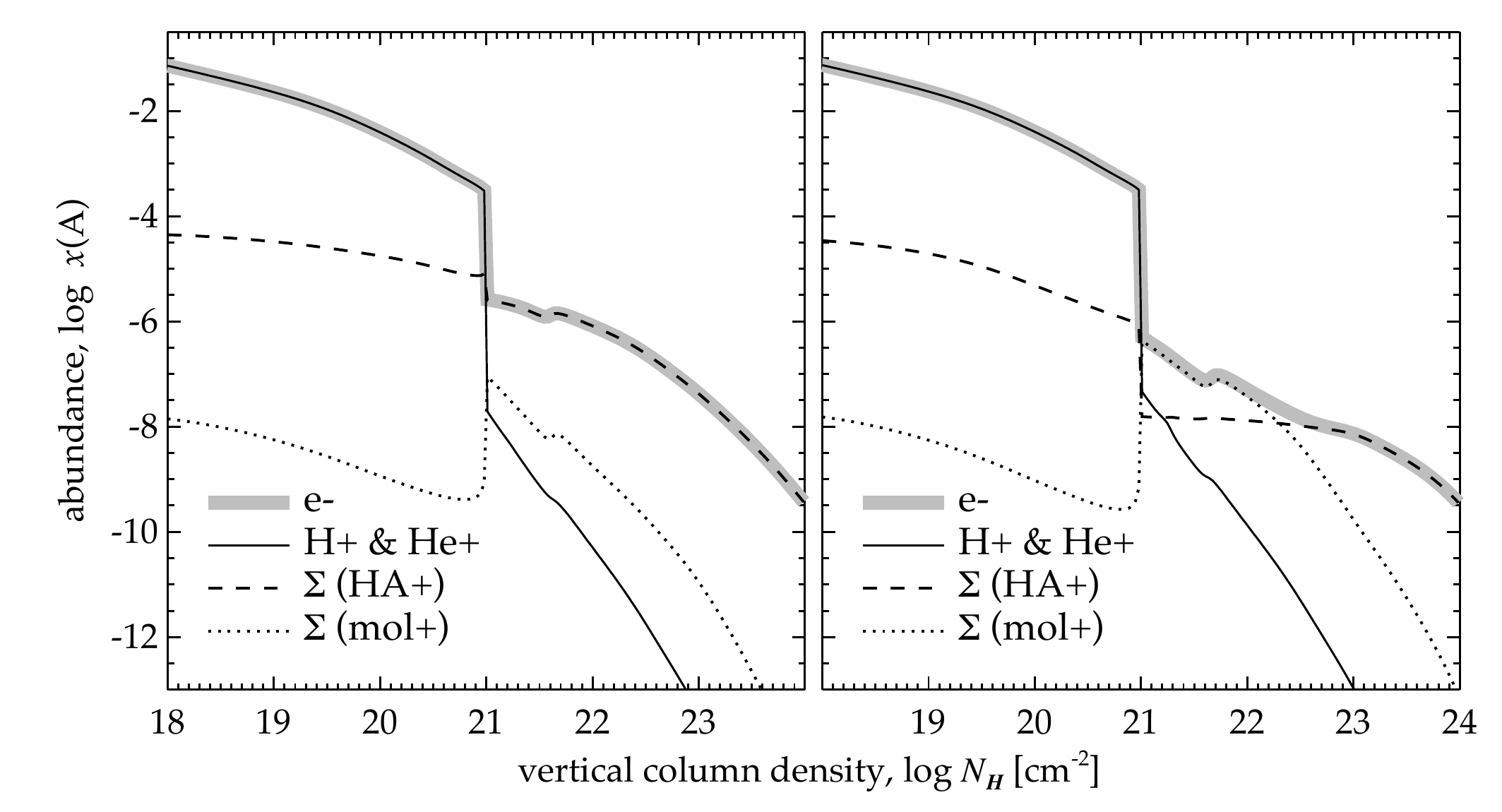}
\caption{Classes of ionic species plotted vs.~vertical column density, $\Nh$,  at 1\,AU.
The three classes are light ions ($\hp$ and $\hep$); the sum of all the heavy 
atomic ions; and the sum of all the molecular ions. The left panel is for the 
depleted interstellar abundances in Table 4; the right panel is for the more
severe depletion in Woodall et al. (2006).}
\end{center} \end{figure}


\begin{references}

\reference{} Anicich, V. G. 1993, J. Phys. Chem. Ref. Data, 22, 1469

\reference{} Aresu, G., Kamp, I., Meijerink, R., Woitke, P., Thi, W.-F. \& 
Spaans, M. 2011, \aap, 526, 163 

\reference{} Badnell, N. 2006,  \apjs, 167, 343

\reference{} Balbus, S. A. \& Hawley, J. F. 1991, \apj, 376, 214

\reference{} Bergin, E., Calvet, N., D'Alessio \& Herczeg, G. J. 2003, 
\apj, 591, L159

\reference{} Chen, D., Gao, H. \& Kwong, V. H. S. 2003, Phys. Rev. A, 
68, 052703 

\reference{} D'Alessio, P., Calvet, N., Hartmann, L., Lizano, S., \& Cant\'o, J. 1999,
 \apj, 527, 893

\reference{} Dalgarno, A., Yan, M. \& Liu, W.-H. 1999, \apjs, 36, 167  

\reference{} Ercolano, B., Drake, J. J., Raymond, J. C. \& Clarke, C. J. 
2008, \apj, 688, 398

\reference{} Ercolano, B., Drake, J. J. \& Clarke, C. J. 2009, \aap, 496, 725

\reference{} Ercolano, B. \& Owen, J. E. 2010, \mnras, 406, 1553

\reference{}  Feigelson. E. D. \& Montmerle, T. 1999, \araa, 37, 363

\reference{} Furlan, E. et al. 2006, \apjs, 165, 568 

\reference{} Furlanetto, S. R. \& Stoever, S. J. 2010, \mnras, 404, 1869

\reference{} Gammie, C. F. 1996, \apj, 437, 353

\reference{} Gao, H. \& Kwong, V. H. S. 2003, Phys. Rev. A, 
68, 052704 

\reference{} Gargaud, M., Hanssen, J., McCarroll, R. \& Valiron, P.
2003, J. Phys. B, 14, 2259

\reference{} Glassgold, A. E., Feigelson. E. D. \& Montmerle, T. 2000, in 
{\it Protostars and Planets IV}, eds. V. Mannings, A. P. Boss, \& S. S. Russell,
(Tucson: Univ. Arizona Press), 429

\reference{} Glassgold, A. E., Najita, J. \& Igea, J. 2004, \apj, 615, 972 (GNI04)

\reference{} Glassgold, A. E., Najita, J. \& Igea, J. 2007, \apj, 656, 515 (GNI07) 

\reference{} Glassgold, A. E., Meijerink, R. \& Najita, J. R. 2009, 
\apj, 701, 142

\reference{} Gorti, U. \& Hollenbach, D. 2004, \apj, 613, 424 

\reference{} Gorti, U. \& Hollenbach, D. 2008, \apj, 683, 287 

\reference{} Guedel. M. et al.~2010, \aap, 519, 2010

\reference{} Hitchcock, A. P. et al. 1988, Phys. Rev. A37, 2488

\reference{} Hollenbach, D. \& Gorti, U. 2009, \apj, 703, 1203

\reference{} Hollenbach, D. \& McKee, C.~F. 1989, \apj, 342, 306

\reference{} Hollenbach, D.~J. \& Tielens, A.~G.~G.~M. 1999, Rev. Mod. Phys., 71, 173

\reference{} Igea, J. \& Glassold, A. E. 1999, \apj, 518, 848

\reference{} Ilgner, M. \& Nelson, R. P. 2006, \aap, 445, 205 

\reference{} Jenkins, E. B. 2009, \apj, 700, 1200
 
\reference{} Kaastra, J. S. \& Mewe, R. 1993, \aaps, 1993, 97, 443

\reference{} Kallman, T. R. \& Palmeri, P. 2007, Rev.~Mod.~Phys., 79, 79

\reference{} Kamp, I., Tilling, I., Woitke, P., Thi, W.-F., \& Hogerheijde, M.
	 2010, \aap, 510, 18

\reference{} Landini, M. \& Monsignori-Fossi, B. C. 1991, \aaps, 91, 183

\reference{} Le Teuff, Y. H., Millar, T. J. \& Markwick, A. J. 2000, \aaps, 146, 167

\reference{} Maloney, P. R., Hollenbach, D. J. \& Tielens, A. G. G. M. 1996, \apj, 466, 561 

\reference{} Markwick, A. J., Ilgner, M., Millar, T. J. \& Henning, Th. \aap, 
385, 632

\reference{} Mattioli, M. et al. 2007 J. Phys. B, 40, 3569

\reference{} Meijerink, R. \& Spaans, M. 2005, \aap, 436, 397

\reference{} Meijerink, R., Glassgold, A. E., Najita, J. R. 2008 \apj, 676, 518 

\reference{} Najita, J. R., \'Ad\'amkovics, M. \& Glassgold, A. E. 2011, in preparation

\reference{} Neufeld, D. \& Dalgarno, A. 1987, Phys. Rev. A. 35, 3142 

\reference{} Opal, C. B., Beaty, E. C., \& Peterson, W. K. 1972, Atomic Data, 4, 209

\reference{} Opal, C. B., Peterson, W. K. \&  Beaty, E. C. 1971, J. Chem. Phys., 55, 4100

\reference{} Oppenheimer, M. \& Dalgarno, A. 1974, \apj, 192, 29

\reference{} Piancastelli, M. N., Hempelmann, F., Heiser, F., Gessner, A., 
R\''udel \& Becker, U. 199, Phys. Rev. A, 59, 300

\reference{} Pneuman, G. W. \& Mitchell, T. P. 1965, Icarus, 4, 494 

\reference{} Preibisch, T. \& Feigelson, E. D. 2005, \aaps, 160, 390

\reference{} Savage, B. D. \& Sembach, K. R. 1996, \araa, 34, 279

\reference{} Shang, H., Glassgold, A. E. \& Shu, F. H. 2002, \apj, 564, 853

\reference{} Shang, H., Glassgold, A. E., Lin, W.-C. \& Liu, C.-F. 2009, \apj, 714, 1733

\reference{} Shapiro, P. R. \& Bahcall, J. N. 1981, \apj, 245, 335

\reference{} Spitzer, Jr., L. 1978, {\it Physical Processs in the Interstellar Medium} (Wiley) 

\reference{} Tarawa, H. \& Kato, T. 1987, At. Data Nuc. Data Tables,
36, 167

\reference{} Turner, N. J. \& Drake, J. F. 2009, \apj, 703, 2152

\reference{} Umebayashi, T. 1983, Prog. Theor. Phys. 69, 480

\reference{} Umebayashi, T. \& Nakano, T. 1988, Prog. Theor. Phys. Supp. 96, 151

\reference{} Verner, D. A. \& Yakolev, D. G. 1995, \aaps, 109, 125 

\reference{} Voronov, G. S. 1997, ADNDT, 65, 1

\reference{} Walsh, C., Millar, T. J. \& Nomura, H. 2010, \apj, 722, 1607

\reference{} Watson, W. D. \& Kunz, A. B. 1975, \apj, 201, 165

\reference{} Watson, W. D. 1976, \apj, 204, 47

\reference{} Weisheit, J. C. \& Dalgarno, A. 1972, Ap. Letters, 12, 103

\reference{} Weisheit, J. C, 1973, \apj, 185, 877

\reference{} Weisheit, J. C. 1974, \apj, 190, 735 

\reference{} Willacy, K. \& Langer, W. D. 2000, \apj, 903, 920

\reference{} Willacy, K., Klahr, H. H., Millar, T. J. \& Henning, Th. 1998, 
\aap, 338, 933

\reference{} Woitke, P., Kamp, I. \& Thi, W.-F. 2009, \aap, 501, 383

\reference{} Woodall, J., Ag\'undez, M., Markwick-Kemper, A. J. \&
Millar, T. J. 2007, \aap, 466, 1197 

\end{references}
\end{document}